\documentclass[12pt,preprint]{aastex}

\def\CIVdblt{{\rm C}\kern 0.1em{\sc iv}~$\lambda\lambda 1548, 1550$}
\def\MgIIdblt{{\rm Mg}\kern 0.1em{\sc ii}~$\lambda\lambda 2796, 2803$}
\def\NVdblt{{\rm N}\kern 0.1em{\sc v}~$\lambda\lambda 1238, 1242$}  
\def\OVIdblt{{\rm O}\kern 0.1em{\sc vi}~$\lambda\lambda 1031, 1037$} 
\def\SiIVdblt{{\rm Si}\kern 0.1em{\sc iv}~$\lambda\lambda1394, 1403$}  

\def\AlII{\hbox{{\rm Al}\kern 0.1em{\sc ii}}}
\def\AlIII{{\hbox{\rm Al}\kern 0.1em{\sc iii}}}
\def\CaII{\hbox{{\rm Ca}\kern 0.1em{\sc ii}}}
\def\CrII{\hbox{{\rm Cr}\kern 0.1em{\sc ii}}}
\def\CII{\hbox{{\rm C}\kern 0.1em{\sc ii}}}
\def\CIII{\hbox{{\rm C}\kern 0.1em{\sc iii}}}
\def\CIV{\hbox{{\rm C}\kern 0.1em{\sc iv}}}
\def\CV{\hbox{{\rm C}\kern 0.1em{\sc v}}}
\def\HI{\hbox{{\rm H}\kern 0.1em{\sc i}}}
\def\HII{\hbox{{\rm H}\kern 0.1em{\sc ii}}}
\def\Lya{\hbox{{\rm Ly}\kern 0.1em$\alpha$}}
\def\Lyb{\hbox{{\rm Ly}\kern 0.1em$\beta$}}
\def\Lyg{\hbox{{\rm Ly}\kern 0.1em$\gamma$}}
\def\Lyfive{\hbox{{\rm Ly}\kern 0.1em$5$}}
\def\Lysix{\hbox{{\rm Ly}\kern 0.1em$6$}}
\def\Lyseven{\hbox{{\rm Ly}\kern 0.1em$7$}}
\def\Lyeight{\hbox{{\rm Ly}\kern 0.1em$8$}}
\def\Lynine{\hbox{{\rm Ly}\kern 0.1em$9$}}
\def\Lyten{\hbox{{\rm Ly}\kern 0.1em$10$}}
\def\HeI{\hbox{{\rm He}\kern 0.1em{\sc i}}}
\def\HeII{\hbox{{\rm He}\kern 0.1em{\sc ii}}}
\def\FeI{\hbox{{\rm Fe}\kern 0.1em{\sc i}}}
\def\FeII{\hbox{{\rm Fe}\kern 0.1em{\sc ii}}}
\def\FeIII{\hbox{{\rm Fe}\kern 0.1em{\sc iii}}}
\def\MnII{\hbox{{\rm Mn}\kern 0.1em{\sc ii}}}
\def\MgI{\hbox{{\rm Mg}\kern 0.1em{\sc i}}}
\def\MgII{\hbox{{\rm Mg}\kern 0.1em{\sc ii}}}
\def\MgIII{\hbox{{\rm Mg}\kern 0.1em{\sc iii}}}
\def\MgIV{\hbox{{\rm Mg}\kern 0.1em{\sc iv}}}
\def\NaI{\hbox{{\rm Na}\kern 0.1em{\sc i}}}
\def\NV{\hbox{{\rm N}\kern 0.1em{\sc v}}}
\def\NII{\hbox{{\rm N}\kern 0.1em{\sc ii}}}
\def\NIII{\hbox{{\rm N}\kern 0.1em{\sc iii}}}
\def\OVI{\hbox{{\rm O}\kern 0.1em{\sc vi}}}
\def\OII{\hbox{[{\rm O}\kern 0.1em{\sc ii}]}}
\def\SiII{\hbox{{\rm Si}\kern 0.1em{\sc ii}}}
\def\SiIII{\hbox{{\rm Si}\kern 0.1em{\sc iii}}}
\def\SiIV{\hbox{{\rm Si}\kern 0.1em{\sc iv}}}
\def\SII{\hbox{{\rm S}\kern 0.1em{\sc ii}}}
\def\SIII{\hbox{{\rm S}\kern 0.1em{\sc iii}}}
\def\SIV{\hbox{{\rm S}\kern 0.1em{\sc iv}}}
\def\TiII{\hbox{{\rm Ti}\kern 0.1em{\sc ii}}}
\def\ZnII{\hbox{{\rm Zn}\kern 0.1em{\sc ii}}}
\newcommand{\kms}{\hbox{km~s$^{-1}$}}
\newcommand{\cmsq}{\hbox{cm$^{-2}$}}
\newcommand{\cc}{\hbox{cm$^{-3}$}}
\def\kms{\hbox{km~s$^{-1}$}}       
\def\cm2{\hbox{cm$^{-2}$}}
\def\cc{\hbox{cm$^{-3}$}}
\def\etal{et~al.\ }

\begin{document}

\title{A Survey of Analogs to Weak {\MgII} Absorbers in the Present \altaffilmark1}

\author{Anand~Narayanan\altaffilmark{2}, Jane~C.~Charlton\altaffilmark{2}, Joe~R.~Masiero\altaffilmark{2,3}, and Ryan Lynch\altaffilmark{2}}
  
\altaffiltext{1}{Based in part on observations obtained with the
NASA/ESA {\it Hubble Space Telescope}, which is operated by the STScI
for the Association of Universities for Research in Astronomy, Inc.,
under NASA contract NAS 5-26555.}
\altaffiltext{2}{Department of Astronomy and Astrophysics, The Pennsylvania State University, University Park, PA 16802, {\it anand, charlton, masiero, rlynch@astro.psu.edu}}
\altaffiltext{3}{Now at Institute for Astronomy, Univ. of Hawaii, 2680 Woodlawn Dr.,
Honolulu, HI, 96822}

\begin{abstract}

We present the results of a survey of the analogs of weak {\MgII}
absorbers (rest--frame equivalent width $W_r(2796) < 0.3$ {\AA}) at
$0 < z < 0.3$.  Our sample consisted of $25$ {\it HST}/STIS echelle
quasar spectra ($R=45,000$) which covered {\SiII}~1260 and {\CII}~1335
over this redshift range.  Using those similar transitions as tracers
of {\MgII} facilitates a much larger survey, covering a redshift
pathlength of $g(z)=5.3$ for an equivalent width limit of {\MgII}
corresponding to $W_r(2796) >0.02$~{\AA}, with $30\%$ completeness
for the weakest lines.
Correcting for incompleteness, we find the number of weak {\MgII}
absorber analogs with $0.02 < W_r(2796) < 0.3$~{\AA} to be $dN/dz = 1.00
\pm 0.20$ for $0 < z < 0.3$.  This compares to a value of $dN/dz =
1.74\pm0.10$ found by \citet{weak1} for the higher redshift range, 
$0.4 < z < 1.4$, and is consistent with cosmological evolution of the
population.  We consider the expected effect on observability of weak {\MgII}
absorbers of the decreasing intensity of the extragalactic background
radiation field from $z\sim1$ to $z\sim0$. Assuming that all the
objects that produce absorption at $z\sim1$ are stable on a cosmological
timescale, and that no new objects are created, we would expect
$dN/dz\sim2$--$3$ at $z\sim0$.  About $30$--$50$\% 
of this $z\sim0$ population
would be decendants of the parsec--scale structures that produce
single--cloud, weak {\MgII} absorbers at $z\sim1$.  The other $50$--$70$\%
would be lower density, kiloparsec--scale structures that produce
{\CIV} absorption, but not detectable low ionization absorption,
at $z\sim1$.  We conclude that at least one, and perhaps some fraction
of both, of these populations has evolved away since $z\sim1$, in order
to match the $z\sim0$ $dN/dz$ measured in our survey.  This would
follow naturally for a population of transient structures whose
generation is related to star--forming processes, whose rate has
decreased since $z\sim1$.

\end{abstract}
\section{INTRODUCTION}
\label{sec:1}

Quasar absorption line systems have been observed at redshifts
$0 < z < 6$.  They provide detailed information about the chemical
content, kinematics, and ionization state of a variety of gaseous
structures ranging from galaxies to the diffuse interstellar medium.
The population of weak {\MgII} absorbers [those with rest frame
equivalent width $W_r(2796) < 0.3$~{\AA}] provides an opportunity
to probe an otherwise elusive part of the network of structures.
\citet{weak2} suggested that they may be produced by an early
population of star clusters or by supernovae ejecta (on large
or small scales) relating to an abundant population of otherwise
invisible galaxies.  Regardless of their mechanism/s of origin, weak
{\MgII} absorbers are likely to be important in a global picture of
star formation, galaxy formation, and/or the interplay between
galaxies and their surroundings.

Because of the practical consideration of simplicity of optical
observations, most of our information about weak {\MgII} absorbers is
for redshifts in the range $0.4 < z < 1.4$.  \citet{weak1}
conducted a comprehensive survey over this range, using high
resolution spectrograph (HIRES; \citep{vogt}) on the Keck I telescope.
They found a redshift path density of $dN/dz = 1.74\pm0.10$ for
weak {\MgII} absorbers with $0.02 < W_r(2796) < 0.3$~{\AA}.  Two
thirds of these absorbers have single clouds, unresolved at a
resolution of $\sim 6.6$~{\kms}, while the others have multiple clouds
spread over tens to hundreds of {\kms}.  Both statistically
\citep{weak1,weak2} and in individual cases
\citep{archive2}, it is found that weak {\MgII} systems arise in
sub-Lyman limit systems [$15.8 < \log N({\HI}) < 16.8$ (cm$^{-2}$)].  In
fact, they comprise a significant fraction, at least 25\%, 
of {\Lya} forest clouds in that column density range
\citep{weak2}.

The most surprising property of the single--cloud, weak {\MgII}
absorbers is their high metallicities.  The metallicity of the gas
producing the {\MgII} absorption is usually constrained to be at
least $10$\% 
solar when constraints are available \citep{weak2}, but
in several of the cases with the best constraints the values are solar
or even supersolar \citep{weak1634}.  This is surprising in view of
the fact that luminous galaxies are rarely found within a $\sim
50 h^{-1}$~kpc impact parameter of the quasar \citep{weak1}.  In some
cases the ratio of {\FeII} to {\MgII} precludes the possibility of
$\alpha$--enhancement from Type II supernovae, and therefore metals
seem to generated ``{\it in situ}'' rather than ejected from distant
giant galaxies \citep{weak2}.  The observed redshift path density
($dN/dz$) of weak {\MgII} absorbers is twice that of the strong
{\MgII} absorbers which are known to be associated with luminous
galaxies [within $38 h^{-1}(L/L^{*})^{0.15}$~kpc; \citep{steidel95}] and which fully account for
almost all of the cross--section such galaxies are expected to
provide.  We are thus drawn to the conclusion that as much of the sky
is covered by significantly metal--enriched regions, that are not
related to giant galaxies, as is covered by such galaxies.

Several other properties of the single--cloud, weak {\MgII} absorbers
are relevant to an interpretation of their origins.  From
photoionization models, the thickness of the {\MgII} absorbing region
is on the order of $1$--$100$~pcs \citep{weak2,weak1634}.  Centered at
the same velocity as the {\MgII} absorption is {\CIV} absorption which
is constrained to arise in a separate higher--ionization region.  
This constraint generally comes from a consideration of whether the
ionization parameters consistent with the low ionization absorption
can produce the observed {\CIV} absorption.
The two phase structure consists of a thin/small {\MgII} absorbing region
(with density $\sim 0.1$~{\cc}) and a larger (hundreds of pcs to a
couple kpcs), lower density region that produces {\CIV} absorption.
The geometries of, and relationship between, these regions is uncertain,
but flattened geometries are generally more consistent with number
statistics and kinematics of the absorbers \citep{nikola}.

There are a number of such indirect clues as to the nature of weak
{\MgII} absorbers, but not yet any definite answers.  Obviously, the
situation would be simplified if a direct connection was found between
the absorber and a nearby, luminous structure.  Such structures might
be dwarf galaxies, star--forming regions in intergalactic cold dark
matter filaments, or intergalactic star clusters of some sort.  If
such objects are to be detected, they must be at low redshift.
Several absorbers have been detected at $z<0.3$ that, although {\MgII}
is not covered in existing spectra, are almost certainly analogs to
the weak {\MgII} absorber population at $z\sim1$.  In the case of
the $z=0.005260$ absorber toward $3$C $273$, there is a post--starbursting
dwarf galaxy within $80 h^{-1}$~kpc of the line--of--sight, which leads
\citet{stocke04} to the conclusion that winds from that dwarf are
responsible for the absorption.  A similar situation has been suggested
to apply in the case of the $z=0.167$ absorber toward PKS~$0405-123$
\citep{chen00}.  It is difficult, however, to be sure that a nearby galaxy
is, in fact, itself responsible for such absorption, rather than a less
luminous object in the same group or cluster.  Also, it may be important
to make a distinction between single and multiple--cloud weak {\MgII}
absorbers which may have different origins \citep{zonak,1206}.
Finding a larger sample of $0 < z < 0.3$ weak {\MgII} absorber analogs
was one of the motivations for this work.  We report on our systematic
survey of the archive of Space Telescope Imaging Spectrograph (STIS)/
Hubble Space Telescope ({\it HST}) observations of quasars.

Our survey facilitates a formal calculation of the redshift path
density of weak {\MgII} absorber analogs at $z\sim0$.  This number can
be compared with the value at $0.4 < z < 1.4$ to consider the
evolution of the population of objects that produce weak {\MgII}
absorption.  Based upon their properties, it is not clear that these
objects should be stable.  Their two phases of gas are not in pressure
balance \citep{weak2,weak1634} and a large amount of invisible matter 
(baryonic or dark) would be needed to attain gravitational stability
\citep{schaye}.  If we are studying the evolution of a transient
population, we are learning about the evolution of the processes that
created the absorbing structures.  On the other hand, for a population of star clusters
formed at very high redshift, we would expect a stable population of
weak {\MgII} absorbers that followed cosmological evolution governed
by a changing extragalactic background radiation.

This paper begins with a description of the {\it HST}/STIS echelle
data that were used for our survey.  We describe how we are able to
survey a significant pathlength by using the similar {\SiII}~1260 and
{\CII}~1335 transitions as tracers of {\MgIIdblt}.  In
\S~\ref{sec:3.1} we first discuss the individual systems found in
our survey, describing their properties and the results of previous
studies of them by other investigators.  We then present in
\S~\ref{sec:3.2} a formal calculation of the redshift number
density, $dN/dz$, correcting for our survey completeness.  The
expected evolution of the population of weak {\MgII} absorbers, taking
into account the changing extragalactic background radiation is
estimated in \S~\ref{sec:4}, and compared to the observed
value.  In \S~\ref{sec:5} the consequences of this for
stability of the structures that give rise to weak {\MgII} absorption
and for the nature of the processes that produce these structures are
considered.  The final section (\S~\ref{sec:6})
contains a brief summary and conclusions.

\section{SURVEY AND DATA ANALYSIS}
\label{sec:2}

In order to detect weak, metal--line features and to cover most of the key
transitions including the Lyman series lines from potential absorbers
at $z < 0.3$, high resolution spectra in the UV are required. The
archived echelle spectra from {\it HST}/STIS were used for our survey.

\subsection{STIS Echelle Spectra}
\label{sec:2.1}

A direct search for {\MgIIdblt} at $z<0.3$ would rely on spectra
obtained with the E230M grating.  As of July 2004, there were about
$15$ spectra available that covered $2800$~{\AA} to $3110$~{\AA}, the
reddest coverage with this grating.  Unfortunately, each spectrum
would only offer a limited redshift coverage of $0 \leq z \leq 0.11$.
Therefore, the maximum cumulative path length for the interval $0 \leq
z \leq 0.11$ is then approximately $15 \times 0.11 = 1.65$, and this assumes
the whole path length exceeds the required equivalent width limit.  We
would expect a maximum of three weak {\MgII} absorbers in this
pathlength if $dN/dz = 1.74$ as at $<z>=0.9$.  Furthermore, the E230M
spectra are quite noisy in the redward regions that would be relevant
to our survey.  This led us to consider and adopt an indirect
approach, using {\SiII}~1260 and {\CII}~1335 as tracers of weak
{\MgII} absorbers in the E140M spectra.

The E140M echelle grating of STIS has a resolving power of $R \sim
45,800$ (corresponding to $6.6$~{\kms}) and simultaneous wavelength
coverage from $1123$ to $1710$~{\AA}.  For our survey we searched $20$
spectra of quasars from the STIS archive that were available before
July 2004. To allow detection of weak metal lines, our search was
limited to spectra which have $S/N
\geq 5$ per pixel over a large fraction of the wavelength
coverage. Table \ref{tab:tab1} lists the $20$ quasar lines of sight
that we searched, with quasar redshifts and some specifics of the {\it
HST}/STIS observations.  The {\MgIIdblt} doublet is not covered for
$z<0.3$ in the many high--quality E140M echelle spectra available in
the archive. To circumvent this limitation we instead used two other
ions, {\SiII} and {\CII}, to trace low ionization absorption.  The
1260~{\AA} transition of {\SiII} and the 1335~{\AA} transition of
{\CII} are moderately strong in absorption and they conveniently fall
within the wavelength regime of the available E140M data set for
redshift $0 < z < 0.3$.  The maximum possible cumulative path length
for our survey would then be $g(z) = 6$, a factor of $3.6$ times
larger than a direct {\MgII} survey using E230M spectra.

The $20$ E140M spectra used for our survey were obtained for a variety
of reasons, and some of these reasons could introduce a bias in our
survey.  For example, a line of sight observed in order to study
particular strong {\Lya} forest absorbers (which were already known to
be metal line systems) could easily be biased toward having a weak
{\MgII} absorber analog.  A line of sight observed for a high velocity
cloud study, however, would have no bias for presence or absence of
weak {\MgII} absorber analogs.  We list in Table \ref{tab:tab1} the
PI of the program to observe each quasar and the program ID to 
aid in consideration of this issue, which we discuss again in
\S~\ref{sec:3.2}.

The reduction and calibration of the E140M spectra were performed
using the standard STIS pipeline \citep{stis1}.  Combination of
separate exposures employed weighting by exposure time.  Similarly,
overlapping orders were combined with equal weight.  Weighting by
inverse variance (as employed, e.g., by \citet{tripp01}) was considered,
but was found to bias the flux downward due to the lower variance of
pixels with smaller counts.  Even smoothing over several pixels would
not eliminate this effect in cases where individual exposures have
$S/N \sim 1$.

For some quasars that were observed at times separated by months or
years, there was often a small shift in the echelle angle, and thus
a change in the wavelength corresponding to a given pixel.  In this case we
selected one scale and combined the others by choosing the nearest
pixel, rather than interpolation which would lead to smoothing.  The
drawback to the procedure we employed is a small, effective decrease in
resolution.  The continuum fit was performed interactively using the
IRAF SFIT task, and employing standard methods as described in
\citet{semsav92}

\subsection{Survey Method}
\label{sec:2.2}

To facilitate comparison to the intermediate redshift survey of
\citet{weak1}, we estimate the rest--frame equivalent width that a
weak {\SiII}~1260 and {\CII}~1335 absorption feature would have,
corresponding to the equivalent width threshold value of 0.3~{\AA} for
{\MgII}~2796. Given the fact that the ionization potentials of Mg, Si
and C are similar, the relative equivalent widths of {\SiII}, {\CII}
and {\MgII} in any system would be largely governed by abundance
pattern and by the oscillator strength of individual electronic
transitions for each atom or ion.  The lack of broad spectral coverage
at high resolving power limits the number of weak systems for which
the equivalent widths of {\SiII}~1260, {\CII}~1335, and {\MgII}~2796
are measured. However, the three single--cloud weak {\MgII} absorbers
found in the spectra of PG~$1634+706$ have all three absorption
transitions covered and detected \citep{weak1634}. The {\SiII}~1260 is
a blended feature in one of the three absorption systems ($z =
0.9055$) and hence only an upper limit could be established for that
particular line. The ratio $W_r(1260)/W_r(2796)$ is $0.67\pm0.17$ for
the $z= 0.8181$ system and $0.68\pm0.26$ for the $z = 0.6534$
system. Similarly the ratios of $W_r(1335)/W_r(2796)$ are 1.45$\pm$0.32
($z=0.6534$), $0.57\pm0.12$ ($z=0.8181$), and $0.85\pm0.04$
($z=0.9055$). Based on these known values, we used the average values
of $0.68$ for the {\SiII}/{\MgII} ratio and $0.96$ for {\CII}/{\MgII}.
These values are also representative of those for isolated components
of strong {\MgII} absorbers \citep{z99,1206,jiethesis3}.
In summary, a metal--line absorber in our survey
is classified as ${\it weak}$ if the rest--frame equivalent
width $W_r(1260) < 0.20$~{\AA} and $W_r(1335) < 0.29$~{\AA}.
Similarly, a detection limit of $0.02$~{\AA} in {\MgII}~2796 (as
used by \citet{weak1}) corresponds to $0.013$~{\AA} in {\SiII}~1260
and $0.02$~{\AA} in {\CII}~1335.

\begin{figure*}
\figurenum{1}
\epsscale{0.8}
\plotone{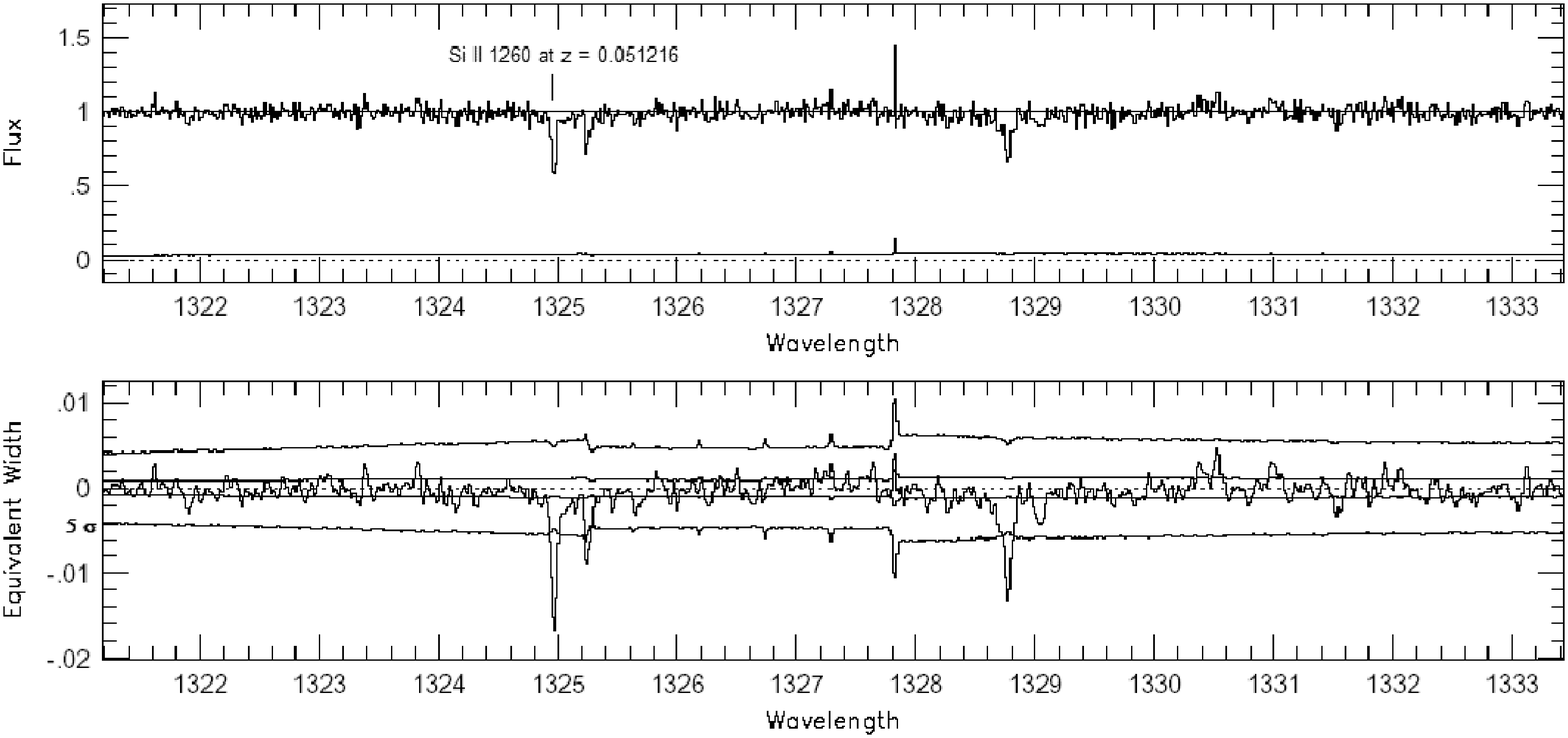}
\protect
\caption{{\SiII} 1260{\AA} detection in the spectrum of PG 1211+143 illustrated from our detection software. The top panel shows the signal spectrum and the uncertainity spectrum. The lower panel shows the equivalent width spectrum. Pixels that show positive equivalent width are emission features and those with negative equivalent width are absorption features. The uncertainity in the equivalent width spectrum is shown at both 1 $\sigma$ (inner) and 5 $\sigma$ (outer) levels. A feature is objectively identified when a pixel has an equivalent width that is larger than the 5 $\sigma$ uncertainity.}
\label{fig:1}
\end{figure*}

To detect weak, low ionization systems we first identified all
absorption features detected at greater than $5\sigma$ significance,
as illustrated in Figure~\ref{fig:1}.
Assuming each identified feature as a redshifted {\SiII}~1260
transition, we searched for a 5$\sigma$ detection at the expected
location of {\CII}~1335. The equivalent width limits for an unresolved
line centered at each pixel were measured, taking into account the
instrumental spread function, following the formalism of
\citet{ss92}. In the event that both {\SiII}~1260 and {\CII}~1335 were
detected at a significance level of 5$\sigma$, we then searched for
absorption in neutral hydrogen, through Ly$\alpha$ and Ly$\beta$
transitions at the derived redshift, in order to confirm the
detection. In the absence of spectral coverage of both these Lyman
series features, we visually inspected the alignment and profile
shapes of the low ionization features of {\SiII} and {\CII}. 

\section{RESULTS FROM SURVEY}
\label{sec:3}

\begin{figure*}
\figurenum{2}
\epsscale{0.4}
\plotone{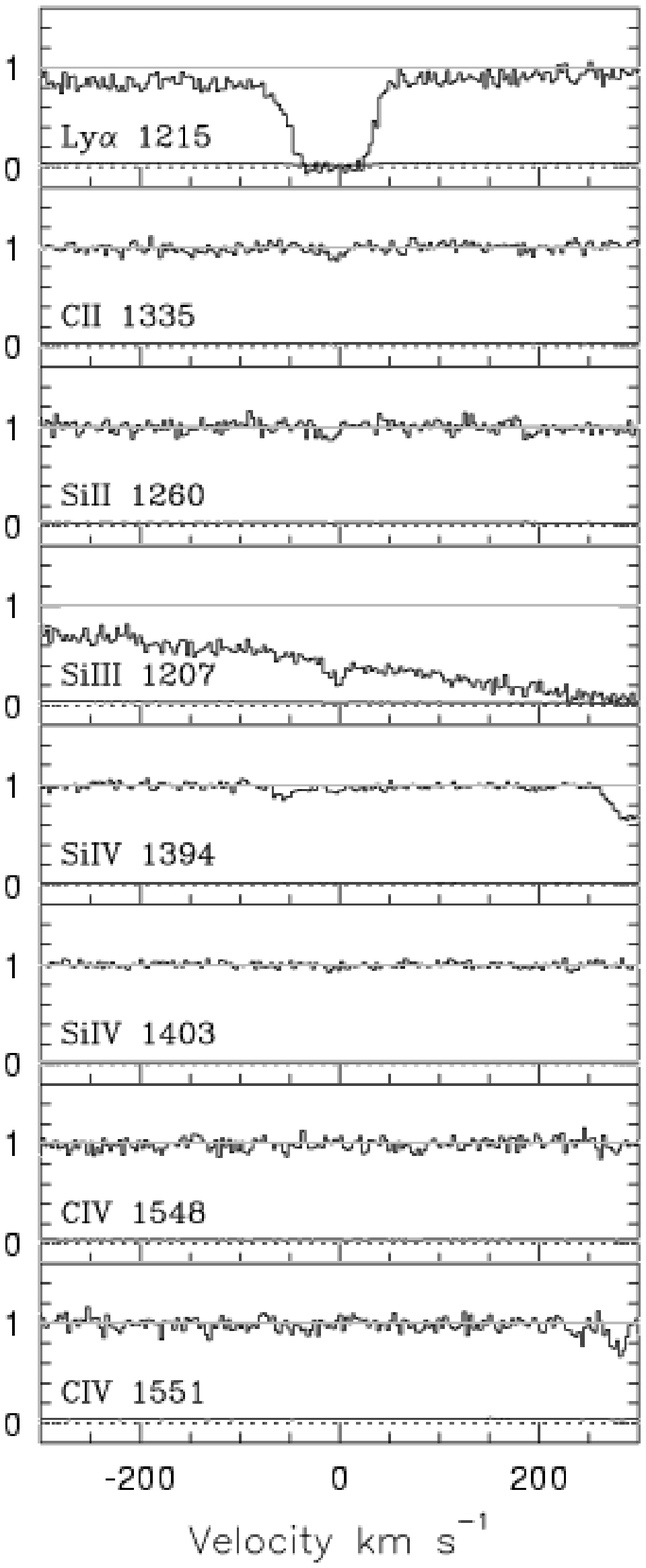}
\protect
\caption{The $z$ = 0.005260 single cloud absorption system found in the 3c273 quasar spectrum, with Voigt profile model fits superimposed for selected transitions. Various key transitions are presented in velocity space. {\SiIII} 1207 is superimposed on Galactic {\Lya} absorption. The {\CIVdblt} is not detected down to a 3 $\sigma$ limit. The absorption system is the closest and weakest low--ionization extragalactic metal line absorber detected so far.}
\label{fig:2}
\end{figure*}

\begin{figure*}
\figurenum{3}
\epsscale{0.4}
\plotone{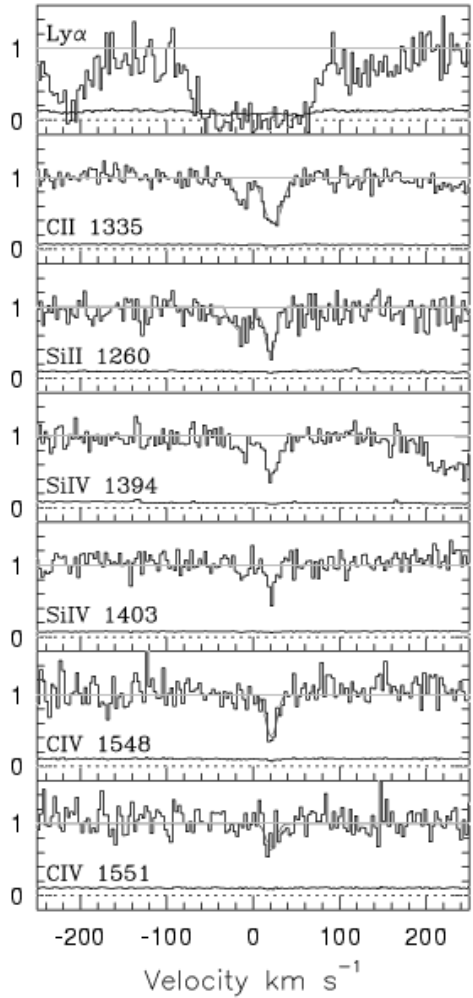}
\protect
\caption{The $z$ = 0.005671 system in the RX~J$1230.8+0115$ quasar spectrum, with Voigt profile model fits superimposed for selected transitions. The transitions are represented in velocity space with zero velocity set at $z$ = 0.005671. This multiple--cloud absorption system is produced by gas in the outskirts of the Virgo cluster.}
\label{fig:3}
\end{figure*}

\begin{figure*}
\figurenum{4}
\epsscale{0.4}
\plotone{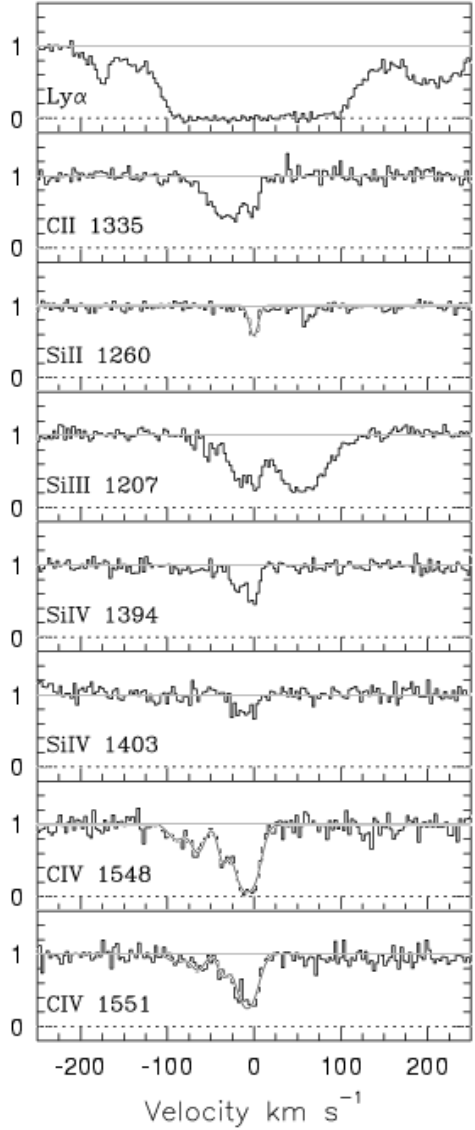}
\protect
 \caption{The $z$ = 0.051216 single--cloud system in the PG 1211+143 quasar spectrum, with Voigt profile model fits superimposed for selected transitions. The zero point of the velocity space is set at $z$ = 0.051216. The blend in the blueward end of {\CII} 1335{\AA} transition feature is identified as absorption from galactic {\SiIV} at 1403{\AA}.}
\label{fig:4}
\end{figure*}

\begin{figure*}
\figurenum{5}
\epsscale{0.6}
\plotone{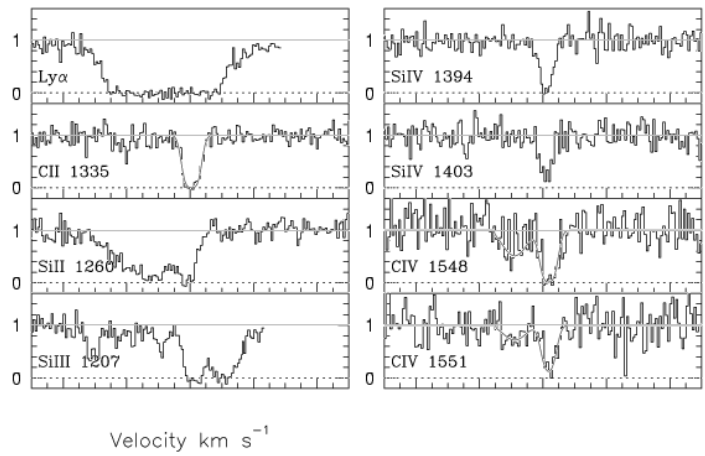}
\protect
\caption{The $z$ = 0.080917 single--cloud system in the PHL 1811 quasar spectrum, with Voigt profile model fits superimposed for selected transitions. The blend in the {\SiII} 1260 and {\SiIII} 1206 absorption featuers could be from metal--poor {\Lya} clouds at redshifts $z$=0.1205 and $z$=0.0735 respectively}
\label{fig:5}
\end{figure*}

\begin{figure*}
\figurenum{6}
\epsscale{0.4}
\plotone{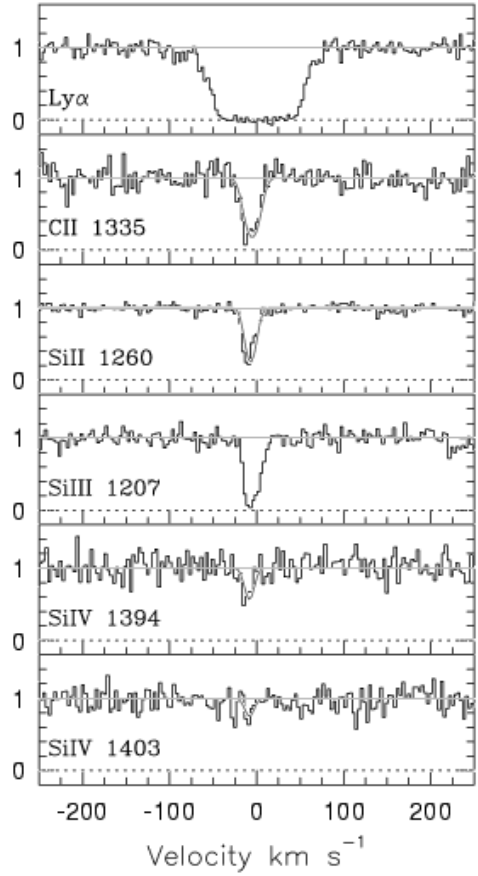}
\protect
\caption{The $z$ = 0.138489 single--cloud system found in the PG 1116+215 quasar spectrum, with Voigt profile model fits superimposed for selected transitions. The {\CIVdblt} is not covered for the STIS E140M spectrum. The absorber could be part of a galaxy group at the system redshift (see \S~\ref{sec:3.1.5})}
\label{fig:6}
\end{figure*}

\begin{figure*}
\figurenum{7}
\epsscale{0.4}
\plotone{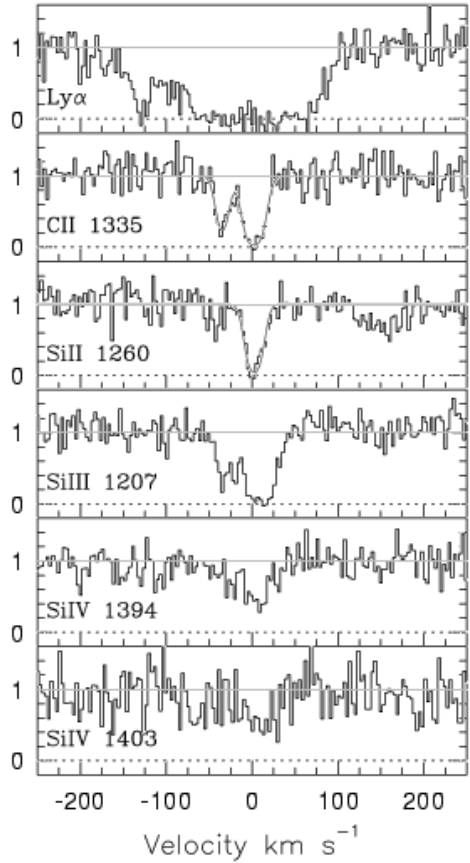}
\protect
\caption{The $z$ = 0.167121 multi--cloud system in the PKS 0405-123 quasar spectrum, with Voigt profile model fits superimposed for selected transitions. The STIS E140M grating does not cover the {\CIVdblt} transition for this sytem. A luminous post--starburst elliptical galaxy has been identified as associated with the absorber (see \S~\ref{sec:3.1.6})}
\label{fig:7}
\end{figure*}

For the $20$ quasar lines of sight, we found six weak {\MgII} analog
absorbers in a redshift pathlength of $\triangle z \sim$ 5.3 within
the redshift interval of 0 $\leq z \leq$ 0.3. Of these, two have
multiple clouds (resolved low ionization spectral features offset in
velocity space) and the other four can be fit as single--cloud,
low ionization absorbers.  The key transitions covered in the
STIS spectra are illustrated for the six systems in
Figures~\ref{fig:2}--\ref{fig:7}, and the equivalent
widths and equivalent width limits for these transitions are
listed in Table~\ref{tab:tab2}. For selected transitions,
the column densities and Doppler parameters derived by 
performing Voigt profile fitting are listed in 
Table~\ref{tab:tab3}. 

\subsection{Discussion of Individual Systems}
\label{sec:3.1}

\subsubsection{3C~273, $z_{abs} = 0.005260$}
\label{sec:3.1.1}
This system is the weakest, low ionization system that we detected
in the STIS E140M archive.  It is also the nearest extragalactic 
metal line system known so far. With $W_r(1260) = 0.010\pm0.002$~{\AA},
it falls below the completeness limit of our survey (see \S~\ref{sec:3.2}).  
This system is unusual for a weak {\MgII} absorber
analog, in that {\CIV}~1548 is not detected to a $3\sigma$ limit
of $W_r(1548)<0.012$~{\AA}.  In a FUSE spectrum, \citet{sembach01}
detect {\Lyb} and six other Lyman series lines and use these to
derive $\log N({\HI})= 15.85$, and possibly detect {\OVI} at a $2-3\sigma$
level.  \citet{tripp02} combine that information from the FUSE spectrum
with the STIS E140M spectrum, and derive physical conditions of the
absorber.  The low ionization gas is found to arise in a surprisingly
thin cloud ($\sim 70$~pc), consistent with sizes derived for some
intermediate redshift single--cloud, weak {\MgII} absorbers
\citep{weak1634}.  The derived metallicity of $[C/H] =-1.2$ is,
however, somewhat lower than that derived for many intermediate
redshift counterparts \citep{weak1634,weak2}. The absorber
is thought to be in the outskirts of the nearby Virgo cluster. 
Specifically, \citet{stocke04} find a dwarf, post--starburst 
galaxy, which is part of the Virgo cluster, at coincident velocity
at an impact parameter of $71 {\rm h}^{-1}$~kpc from the
line of sight, and use this to support the idea that winds
from that galaxy are responsible for the absorption.

\subsubsection{RX~J$1230.8+0115$, $z_{abs} = 0.005671$}
\label{sec:3.1.2}
This system is clearly a multiple--cloud, weak {\MgII} absorber
analog, with two components resolved in {\SiII}~1260 and
{\CII}~1335.  The {\CIV} absorption follows the same kinematics,
and is relatively weak compared to the low ionization absorption.
The {\SiIVdblt} doublet is also detected.  Using the ratios of
these absorption features, \citet{rosenberg} determined that
a single phase, photoionized medium is consistent with the data,
providing a simultaneous fit to transitions of various ionization.
It should be noted, however, that an additional higher density
phase is not excluded by the data.
This quasar offers a different line--of--sight
through the outskirts of the Virgo cluster as it only $350 {\rm h}^{-1}$~kpc 
(less than 1$^o$ in the sky) from the 3C~273 line
of sight. This absorber is also part of the cluster environment
at a coincident velocity with the
absorber discussed previously, at $z_{abs}=0.005260$ along that
line of sight. 
The large line--of--sight thickness of $\sim 20$~kpc that
resulted from this relatively highly ionized cloud ($\log U \sim -2.7$)
is in contrast to the small size of $\sim 70$~pc derived for the
3C~273 absorber \citep{tripp02}.  This led \citet{rosenberg}
to suggest that a large filamentary structure, perhaps related to
a wind, connects the two lines of sight.
This line of sight was selected for observation because of its proximity
to $3C~273$ toward the Virgo cluster.  Because of its special location
there could be a bias introduced to our survey (most likely a
positive one), but no specific absorption system was known to
exist in advance of the observation.

\subsubsection{PG~$1211+143$, $z_{abs} = 0.051216$}
\label{sec:3.1.3}
The absorber is a single--cloud, weak {\MgII} system analog. The
{\CII}~1335 absorption profile is blended with Galactic {\SiIV}~1403.
The blend is confirmed by the detection of {\SiIV}~1394 at $z =0$.
The {\CIVdblt} and {\SiIVdblt} resonant doublets for the absorber are
covered and detected in the spectrum and are likely to be produced by
a separate high ionization phase of the absorber.  Multiple components
are clearly detected in {\CIV}, extending blueward of the strongest
component.  \citet{stocke04} have previously reported the presence of
this metal line absorber. Although current imaging surveys are below
the sensitivity limit to detect a dwarf galaxy ($m_B > 22$) at the
redshift of the absorber, a more luminous galaxy was found at a
projected distance of $\sim 100{\rm h}^{-1}$~kpc from the line of
sight at a comparable recession velocity (\citet{stocke04}, \citet{Tumlinson}).  
Since the PG~$1211+143$ line of sight
was observed in order to study two strong {\Lya} forest absorbers that
were known to have {\SiIII} absorption, it was quite likely to have a weak {\MgII}
absorber analog.

\subsubsection{PHL~$1811$, $z_{abs}=0.08093$}
\label{sec:3.1.4}
We classify this system as a single--cloud, weak {\MgII} absorber
analog, and the strength is directly confirmed by detection of
{\MgIIdblt} with $W_r(2796)=0.145\pm0.053${\AA} in a low resolution,
G230 spectrum from STIS \citep{jenkins03}.  In the E140M spectrum, the
{\SiII}~1260 and {\CII}~1335 transitions are saturated, but can be
consistently fit with a single component. Both {\SiII}~1260 and
{\SiIII}~1206 are blended, possibly with {\Lya} absorption from
metal--poor regions at redshifts $z=0.1205$ and $z=0.0735$,
respectively.  There is {\CIV} centered at the same velocity of the
low ionization component, and an offset component at $\sim -50$~{\kms}
is also possibly detected in a noisy region of spectrum.  The {\Lya}
is also asymmetric relative to the low ionization gas, with additional
absorption to the blue.  This system has been studied by \citet{jenkins03},
using a FUSE spectrum and G140L and G230L spectra from STIS.  In the
FUSE spectrum, {\OVI} is detected, and the Lyman series lines and
Lyman limit break indicates that this is a Lyman limit system, with
$\log N({\HI}) > 17.5$ \citep{jenkins03}.  This is noteworthy, since
most weak {\MgII} systems at $z\sim1$ do not produce Lyman limit
breaks.  There is an $L_*$ galaxy consistent with the redshift of the
absorber at an impact parameter of $34 {\rm h}_{70}^{-1}$~kpc
\citep{jenkins03}. Thus it appears this may be an absorber similar to
a strong {\MgII} absorber, which might tend to produce weak low
ionization absorption because of a relatively high impact parameter.
Since this line of sight was observed in order to study the same
particular system that we have found in our survey, it clearly was biased.

\subsubsection{PG~$1116+215$, $z_{abs} = 0.138489$}
\label{sec:3.1.5}
By our fitting procedure, this system is a single--cloud, weak
{\MgII} analog absorber.  A possible second component did not
significantly improve (by an ``F-test'') a simultaneous fit to
the {\SiII}~1260 and {\SiII}~1193 lines or a fit to the
{\CII}~1335 line.  However, \citet{sembach04} do claim a second
component is needed to fit the low ionization lines, and this is
certainly a possibility especially in view of the similarity of
their line profiles.  In this case, the system is a multiple--cloud,
weak {\MgII} absorber with a very weak offset component close in
velocity to the primary.

The {\CIVdblt} doublet is not covered in the STIS E140 spectrum. However
{\CIV}~1548 is just detected, with $W_r=0.15\pm0.04$~{\AA} in a low
resolution FOS/HST spectrum.  This system is discussed in depth by
\citet{sembach04}, who combine information from this STIS spectrum
with that from a high--quality FUSE spectrum.  The Lyman series lines,
a weak partial Lyman limit break, various low and intermediate
ionization lines, {\NV}~1239, and {\OVI}~1032 were
detected in the FUSE spectrum.  \citet{sembach04} noted that this
absorber had a particularly high ratio of {\HI} to {\OVI}, which
they took as evidence for its multiphase/multi--temperature nature.
No specific galaxy has been identified with this absorber, but there
is an excess of galaxies at this redshift indicating a possible
galaxy group \citep{sembach04,tripp98}.  This line of sight was
proposed for observation in order to study a Milky Way high
velocity cloud, so it is not biased for the purpose of our
survey.

\subsubsection{PKS~$0405-123$, $z_{abs} = 0.167121$}
\label{sec:3.1.6}
This system is a multiple cloud weak {\MgII} analog with two distinct
clouds separated by $\sim40$~{\kms}.  The E140M spectrum does not
provide coverage of {\CIVdblt}. It is however detected in a FOS/HST
spectrum, with $W_r(1548) = 0.54\pm0.05$~{\AA} \citep{kp13}.
Higher ionization transitions, {\OVI} and {\NV}, are also detected in
the STIS spectrum, clearly indicating multiphase conditions \citep{chen00}.
The {\HI} for this system is detected
in the STIS spectrum through Ly$\alpha$ and Ly$\beta$ features. Based
on the derived column density of neutral hydrogen
($N({\HI}) \sim 10^{16}$ cm$^{-2}$), the system is classified as a partial Lyman
limit system \citep{chen00}. \citet{spinrad93} report the
presence of a spiral (L$_K$ = 0.02 L$_K^*$) and a luminous elliptical
galaxy (L$_K$ = 1.12 L$_K^*$) at the absorber's redshift, at
impact parameters of 63 and 75 kpc.  The elliptical galaxy has
spectroscopic signature of a recent star burst event (Spinrad
{\etal}1993).  \citet{chen00} infer a metallicity greater than
$10$\% solar, with at least a solar value being most likely.
Based on these facts, they suggest
weak {\MgII} analog absorber is created by heavier elements
transported from the elliptical galaxy by the star burst activity.
This line of sight is only weakly biased for the purpose of our
survey due to its proximity to Virgo.  The original observation
was for the purpose of {\Lya} forest studies, but no specific
metal-line systems were targetted.

\subsection{Survey Completeness {\&} Redshift Number Density}
\label{sec:3.2}
\begin{figure*}
\figurenum{8}
\epsscale{1.0}
\plotone{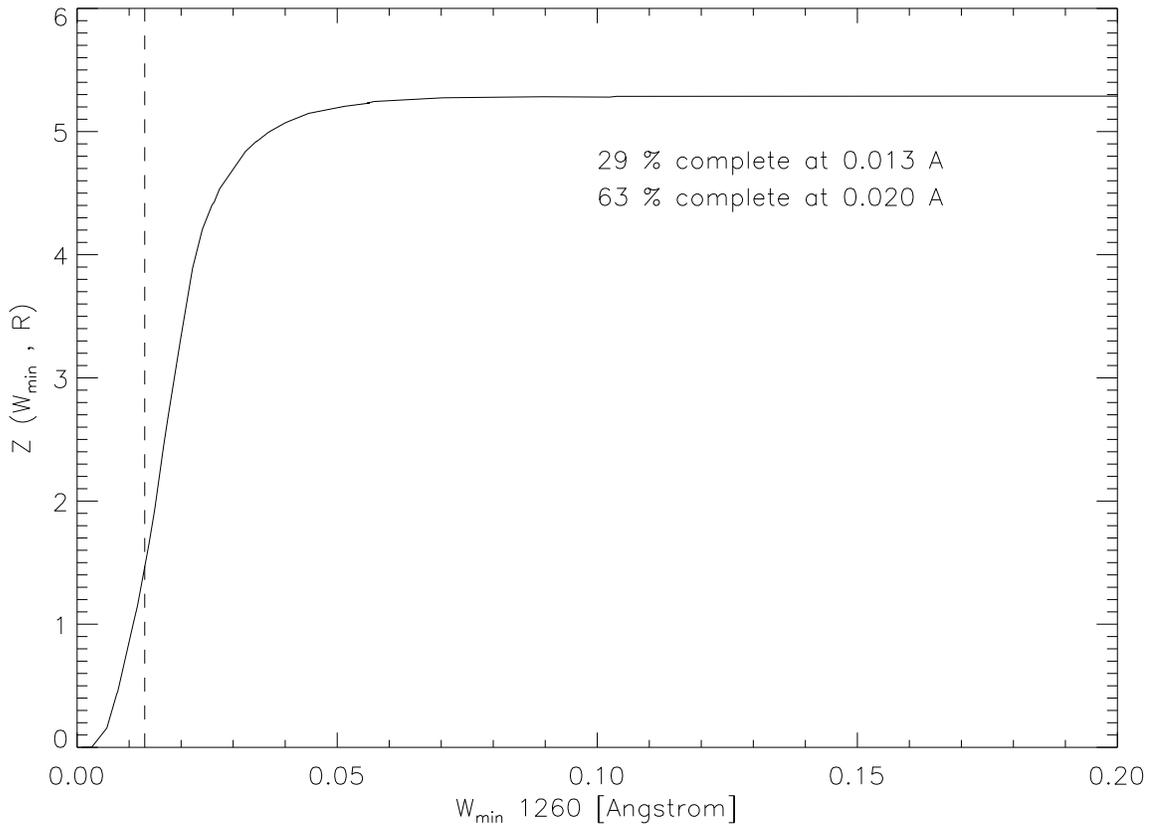}
\protect
\caption{Cumulative redshift path of the survey as a function of {\SiII} 1260 equivalent width for a 5 $\sigma$ detection of both {\SiII} 1260 and {\CII} 1335. The redshift interval is from 0 $\leq z \leq$ 0.3. $R$ is defined as the ratio of the equivalent--widths of {\CII}~1335 and {\SiII}~1260 (see \S~\ref{sec:2.2})}
\label{fig:8}
\end{figure*}

The completeness of the survey as a function of redshift was
calculated by using the formalism given by \citet{ss92}
and \citet{ltw87}.
The cumulative redshift path length covered by the survey of $20$
quasars over the redshift range, $0 < z < 0.3$ is given by
\begin{equation}
Z(W_r,R) = \int_{0}^{0.3}g(W_r,z,R)dz ,
\end{equation}
\noindent where $g$(W$_r$,$z$,$R$) is the function that gives the number
of sight lines along which {\SiII}~1260 and {\CII}~1335, at redshift $z$
and with rest--frame equivalent width greater than or equal to W$_r$ in
{\SiII}~1260, could have been detected at a 5$\sigma$ level. We define
$R$ as the expected ratio of the equivalent--widths of {\CII}~1335 and
{\SiII}~1260.  As discussed in section \S~\ref{sec:2.2}, we
adopt $R=1.41$ based on empirical results.
Table~\ref{tab:tab2} lists the values of cumulative redshift path length over
which the individual systems that we found in our survey could have
been detected, and Figure~\ref{fig:8} presents the completeness
as a function of $W_r(1260)$. Our survey is $30 \%$ complete for an equivalent
width threshold of $W_r(1260) = 0.013$~{\AA} ($R = 1.41$),
corresponding to $W_r(2796) = 0.02$~{\AA}. For comparison, the
\citet{weak1} survey at $z \sim 1$ was $80 \%$ complete
at $W_r(2796) = 0.02$~{\AA}.
Although, our survey is less complete than \citet{weak1} at small
equivalent widths, we note that it is quite complete for an
equivalent width threshold of $W_r(1260)=0.02$~{\AA}, and that
we are nevertheless correcting for incompleteness.
Below we consider the effects of eliminating quasars with low $S/N$ spectra
from the sample.

The number of absorbers per unit redshift
($dN/dz$) was calculated as the sum of the reciprocal of the
cumulative redshift path lengths, using the formalism given in
\citet{ltw87}:
\begin{equation}	
\frac{dN}{dz} = \sum_{i}^{N_{sys}} [Z(W_i,R_i)]^{-1}
\end{equation}
\noindent and the variance in $dN/dz$ is given by
\begin{equation}
\sigma_{dN/dz}^{2} = \sum_{i}^{N_{sys}}[Z(W_i,R_i)]^{-2}
\end{equation}
To facilitate comparison with the \citet{weak1} estimation
of $dN/dz$ at $0.4 \leq z \leq 1.4$, the rest--frame equivalent width
upper and lower limits for {\SiII}~1260 were chosen to be $0.2$~{\AA}
and $0.013$~{\AA}, respectively. Over the redshift range of $0 \leq z \leq 0.3$
five systems were detected for the equivalent--width ranges
of $0.013 \leq W_r(1260) < 0.2$~{\AA} and $0.019 \leq W_r(1335) < 0.29$~{\AA},
and the redshift number density was found to be 
$dN/dz = 1.0 \pm 0.20$.

The system at $z = 0.005260$ along the line of sight towards 3C~273 was
excluded while computing the redshift path length $Z(W_{i},R_{i})$,
since the {\SiII}~1260 was only detected at a $4.5\sigma$ level, and
the rest--frame equivalent width of its {\SiII}~1260 was only
$0.010$~{\AA}, less than the equivalent width lower limit of
$0.013$~{\AA}.  

\begin{figure*}
\figurenum{9}
\epsscale{1.0}
\plotone{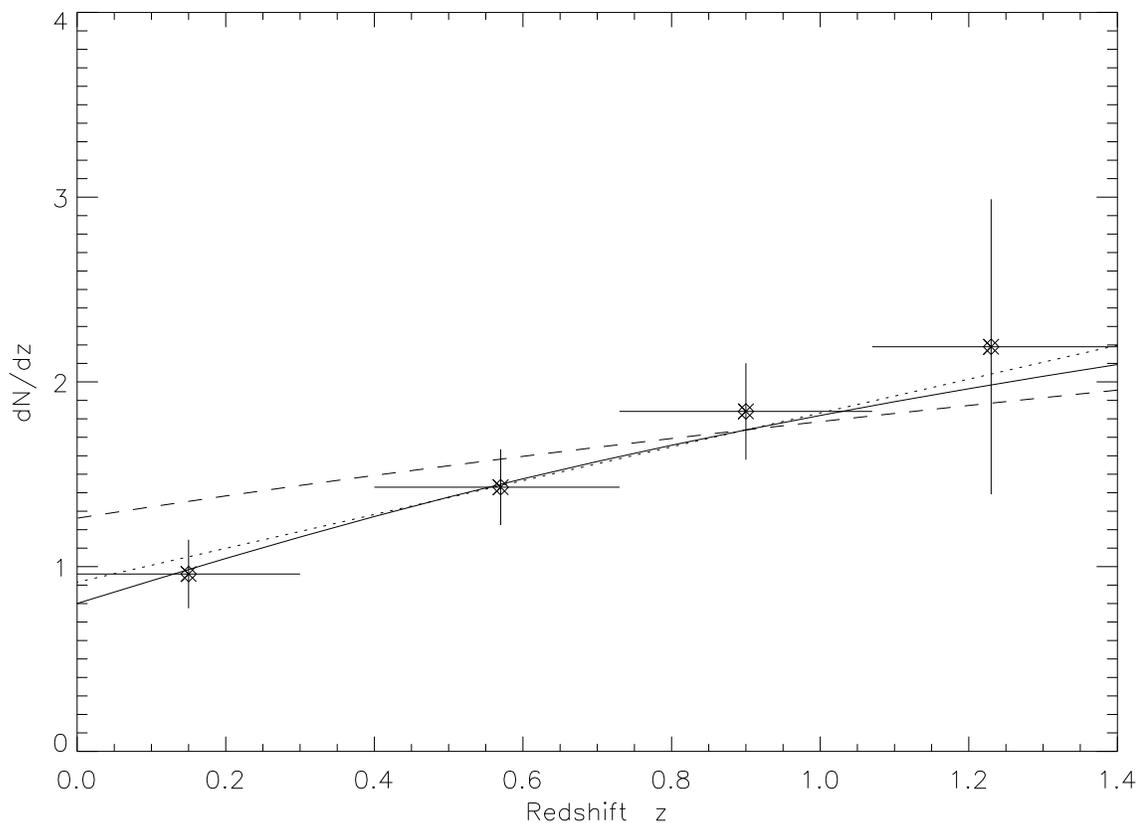}
\protect
\caption{Number of weak {\MgII} absorbers per unit redshift for four redshift bins over the interval 0 $\leq z \leq$ 1.4. Vertical error bars are the Poission uncertainities in the dN/d$z$ values. Horizontal lines are the redshift bins. The dN/d$z$ values for redshifts $z >$ 0.3 are taken from Churchill {\etal}(1999). The solid curve is the no--evolution expectation for a $\Lambda$CDM universe with $\Omega_m$=0.3 and $\Omega_{\Lambda}$=0.7. The dotted curve is the no--evolution expectation for q$_o$ = 0 ($\gamma$ = 1) and the dashed curve is for q$_o$ = 0.5 ($\gamma$ = 0.5). The curves have been normalized at $z$=0.9 and dN/d$z$ = 1.74.}
\label{fig:9}
\end{figure*}

Figure~\ref{fig:9} is a plot of $dN/dz$ versus redshift for the
equivalent width threshold range corresponding to $0.02 \leq W_r(2796)
<0.3$~{\AA}, with our $<z> = 0.15$ point included along with higher
redshift values from \citet{weak1}. The curves represent the
no-evolution expectations for a $\Lambda$CDM universe ($\Omega_m$=0.3 and $\Omega_{\Lambda}$=0.7; solid curve),
a critical universe with deceleration
parameter $q_{0} = 0.5$ (dashed curve), and an open universe with
$q_{0} = 0$ (dotted curve), all normalized to $dN/dz = 1.74$ at $z =0.9$.
The redshift dependence of the number density of absorbers is consistent
with no evolution.

\subsection{Possible Systematics and Biases of $dN/dz$}
\label{sec:3.4}

Two potentially important parameters that could affect our observed
$dN/dz$ value are the ratios $W_r(1260)/W_r(2796)$ and
$W_r(1335)/W_r(2796)$.  As we described in \S~\ref{sec:2.2},
we estimate the typical values of these ratios based upon the
several weak {\MgII} absorbers for which all three transitions
were covered in a spectrum.  There is a spread of values, and
even the mean is not well determined.  However, our method for
determining $dN/dz$ and the particulars of our dataset imply that
the value we would obtain for $dN/dz$ is not sensitive to the
choice of the equivalent ratios, within their possible range.

More specifically, only if the ratio $W_r(1335)/W_r(2796)$ was
taken to be greater than $1.5$ would the system toward PG~$1211+143$
(see Table~\ref{tab:tab2})
be excluded from our survey.  In this unlikely case, there would only be
four terms included in our $dN/dz$ calculation using the reciprocals
of cumulative path lengths and the result would be $dN/dz=0.74\pm0.14$,
not very different from our actual value.

Similarly, only if both $W_r(1335)/W_r(2796)$ and $W_r(1260)/W_r(2796)$
were less than $0.5$, would the system toward 3C~273 be included in
the $dN/dz$ summation.  Based on experience with both strong and
weak {\MgII} systems, it seems unlikely that {\SiII} would be so weak
compared to {\MgII}.  Nonetheless, if we compute $dN/dz$ including
the 3C~273 system, we obtain $dN/dz = 2.0\pm1.2$.  The large error
value is due to the large incompleteness ($\sim18$\%) of the survey
for systems as weak as that one.

We adopted the lower limit of $W_r(1260) \geq 0.013$~{\AA} for our survey
to facilitate comparison to \citet{weak1}.  However, we noted above
that the survey is only $30 \%$ complete at that limit.  We don't expect
any biases due to this incompleteness, and the formal error in our
calculated $dN/dz$ appropriately includes the uncertainty.  Also, for the
weakest system actually included in our survey (as opposed to the
weakest that could have been found), the completeness was 73\%
so that the correction calculated by equation (2) was relatively small.
Nonetheless,
we have done the experiment of eliminating the noisiest $10$ spectra
and recomputing $dN/dz$ for the remaining 10 quasars.  With these
$10$ spectra  our completeness at $W_r(1260) = 0.013$~{\AA} is $\sim60$\%.
For this reduced sample, we estimate $dN/dz = 1.0 \pm 0.6$.
This is not different than our value from the full
sample of $20$ quasars, thus we adopt that value, with its smaller
error, for further discussion.

As discussed for each of the individual weak {\MgII} absorber analogs
in our sample in \S~\ref{sec:3.1}, this survey is biased because
some lines of sight were observed because of known absorbers that were
likely to fall in this category.  To first order, the bias is likely
to yield a larger $dN/dz$ value than an unbiased survey.  Indeed, all
of the weak {\MgII} absorber analogs that we detected in this survey
had already been studied by other authors.  Because of this, we can
consider the number $dN/dz = 1.0 \pm 0.20$ an upper limit.  The
correct value could be somewhat lower.  However, we should note
that there are quasars in the sample that would appear to have a
similar level of positive bias that do not have detected weak {\MgII}
absorber analogs.  We do not think this systematic bias is large,
and, as it turns out, even a factor of $2$ reduction in the value
would not affect our basic conclusions.

After considering these possible systematics and biases, we expect
that our observed value of $dN/dz = 1.0 \pm 0.20$ for $0.02 < W_r(2796) < 0.3$~{\AA} at $0 < z < 0.3$ is an accurate representation of the true
value.  Systematics due to assumptions about the relationship between
the {\SiII}, {\CII}, and {\MgII} are unlikely to be large.
The largest inaccuracy is judged to be a possible bias in
quasars selected for STIS observations toward the detection of
weak {\MgII} absorbers.

\section{EXPECTED EVOLUTION OF ABSORBERS DUE TO CHANGING EXTRAGALACTIC BACKGROUND RADIATION}
\label{sec:4}

The population of structures that produce weak {\MgII} absorption
could be different at different redshifts due to a variety of
factors.  The process of hierarchical structure growth and
merging will lead to a change in the distribution of the total
hydrogen column density in the universe.  The metallicity of
the universe will gradually increase and supernovae and their
cumulative effects, superbubbles and superwinds, will redistribute gas.
Finally, the ionization state of the absorbers will change due to
the evolution of the ionizing extragalactic background radiation (EBR).
Due to the latter effect, even a stable object with an unchanging
total density distribution will be observed to have different
absorption properties at different redshifts.

The EBR is known to change with redshift, because of declining
space density of bright quasars and star forming galaxies from 2 $\geq
z \geq$ 0 \citep{hm96,hmconf}.
Therefore a population of weak {\MgII}
absorber structures in the present epoch is expected to have
different ionization properties and phase structures compared to the
absorber population detected at higher redshifts.  We consider, 
through examples, and statistically, the expected changes to
the populations of structures that exist at $z\sim1$ as they
evolve to the current epoch.  This lends insight into the nature
of the population of weak {\MgII} absorbers that is observed in
the present.

\subsection{Changes to Phase Structure of Weak {\MgII} Absorbers}
\label{sec:4.1}

The \citet{hmconf} spectrum defines the mean specific intensity of the
radiation field as a function of energy and redshift.  It is generated
based on measurements of the QSO luminosity function
and the observed {\Lya} cloud population which absorbs the EBR.  The
Haardt and Madau formalism also offers provision for adding photons
from star forming galaxies using stellar spectral energy distributions
from \citet{bruzualcharlot}.  We adopt a QSO plus star forming galaxy
model with a $10$\% escape fraction for our calculations, but
also consider a ``QSO--only'' model as the opposite extreme.  For both
models, we include the effect of absorption by the inter--galactic
medium (using observed parameters as described in \citet{hm96} and
\citet{hmconf}).  The predictable nature of
the EBR allows us to hypothesize specifically how the phase structure
of the weak {\MgII} absorber population at $z\sim1$ would change under
the influence of a less intense metagalactic flux in the present
epoch.  Qualitatively, the decreasing intensity of the EBR should lead
to an increase in the ratio of low to high ionization states of an
atom (e.g., {\SiII}/{\SiIV} and {\CII}/{\CIV}) for a gaseous
metal--line absorber.  A larger contribution to the EBR by star
forming galaxies will, in general, lead to a lower ionization state
due to the relative absence of high energy photons.  However, none of
our conclusions below are significantly altered by varying assumptions
about escape fraction from star forming galaxies
or about absorption by the {\Lya} forest, as we will mention when
relevant.

\subsection{Analysis of Evolution using Sample Weak {\MgII} Systems}
\label{sec:4.2}

Photoionization calculations using Cloudy \citep{cloudy01} have
yielded estimates of densities, metallicities and
temperatures in the low and high ionization phases of weak
{\MgII} gas clouds at $z \sim 1$ \citep{weak2,weak1634,jiethesis3}.
Almost all of the weak {\MgII} absorbers for which information is
available have been demonstrated to have a multi--phase structure.
The low ionization phase has a higher density and a
smaller thickness/size than the high ionization phase. To guide
our thinking on the evolution of weak {\MgII} absorbers we considered
the expected evolution of three specific single cloud weak {\MgII}
absorbers at redshifts of $z = 0.81$, $z= 0.90$, and $z=0.65$ along
the line of sight to the quasar PG 1634+704 \citep{weak1634}.
These three absorbers appear to have phase structures that are
typical of other weak {\MgII} systems that were found at intermediate
redshifts \citep{weak2}, and can hence be treated as representatives
of the absorber population. For detailed photoionization models for the
three systems, see \citet{weak1634}.

\begin{figure*}
\figurenum{10}
\epsscale{0.8}
\plotone{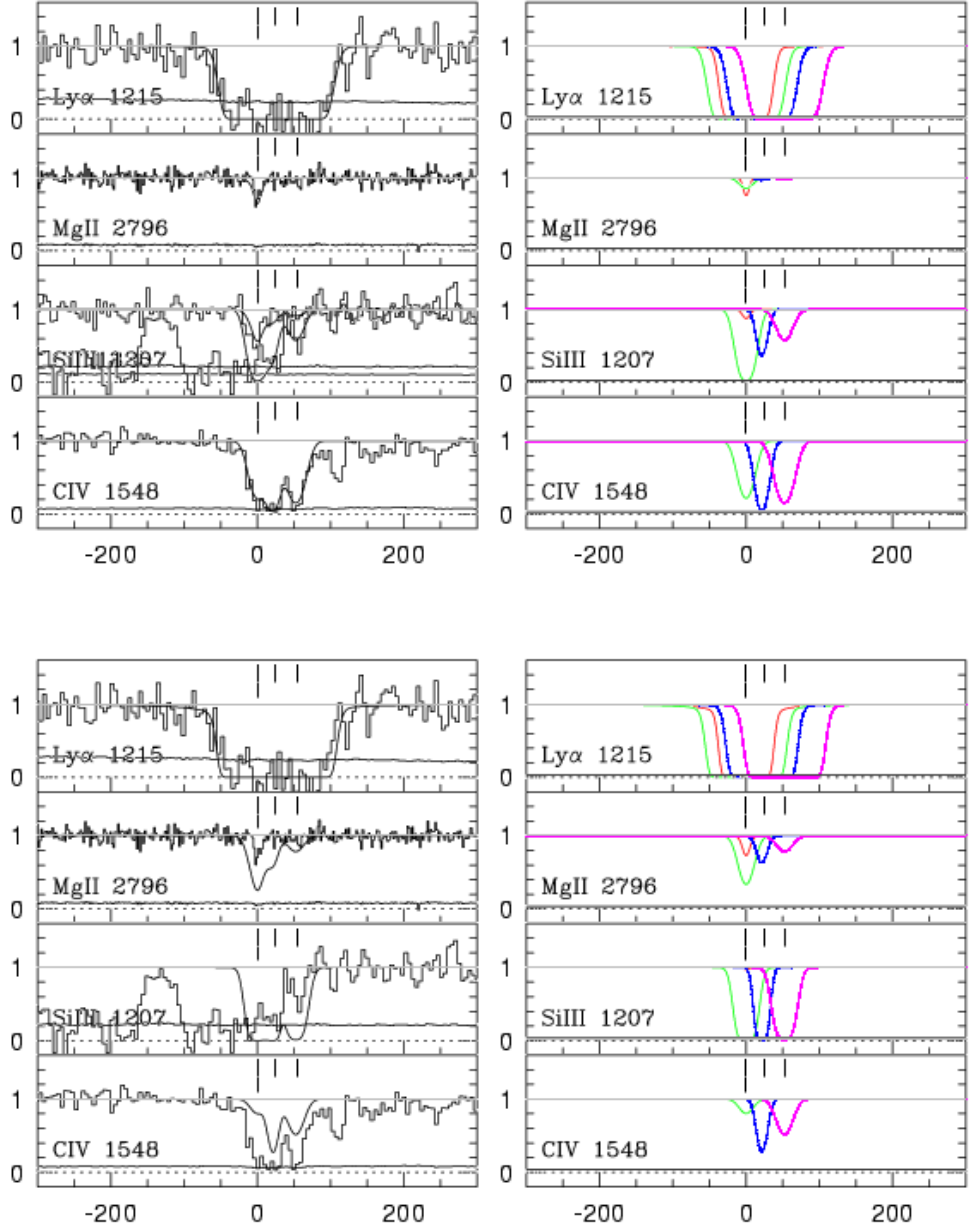}
\protect
\caption{\small{Evolution of the $z \sim$ 0.8 weak {\MgII} absorber from its observed redshift to the present epoch. {\bf Top Left Panel:} Data with photoionization model from Charlton {\etal}(2003) superimposed. {\bf Top Right Panel:} The low density, high ionization and high density, low ionization phases shown separately. The synthetic spectrum shown in {\it red} is the high density {\MgII} phase centered at 0 km/s in velocity space. The {\it green} synthetic spectrum is the contribution from the low density {\CIV} phase. The {\CIV} phase is centered on the {\MgII} phase (an offset of 0 km/s). {\bf Bottom Left Panel:} Shows how the absorption feature of a similar system present at $z$ = 0 would look like. The data are superimposed on the synthetic spectrum to highlight changes in absorption feature strengths for individual transitions. {\bf Bottom Right Panel:} The dense, parsec--sized {\MgII} phase and less dense, kiloparsec--sized {\CIV} phase are shown separate for the $z=0$ scenario. The extent of change in column density for {\MgII}, {\CIV} and {\HI} are summarized in Table~\ref{tab:tab5}}}
\label{fig:10}
\end{figure*}

\begin{figure*}
\figurenum{11}
\epsscale{0.8}
\plotone{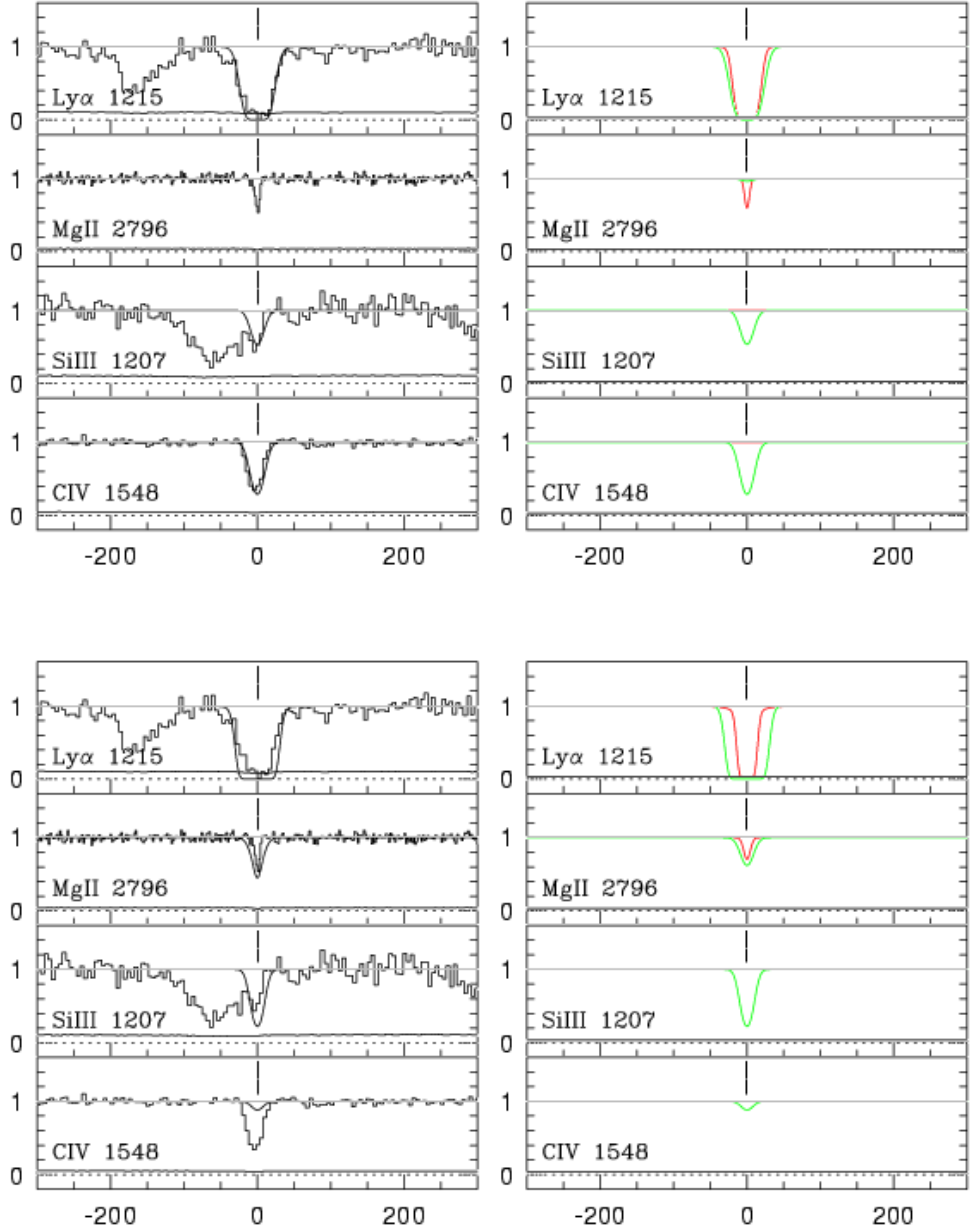}
\protect
\caption{Same as Fig~\ref{fig:10} but showing the evolution of the $z \sim$ 0.9 weak {\MgII} absorber from its observed redshift to the present epoch. In this case, there are two low density phases, one centered on the {\MgII} phase and the other offset by +15 km/s in velocity space. The change in column density for the {\MgII}, {\CIV} and {\HI} are tabulated in Table~\ref{tab:tab6}}
\label{fig:11}
\end{figure*}

\begin{figure*}
\figurenum{12}
\epsscale{0.8}
\plotone{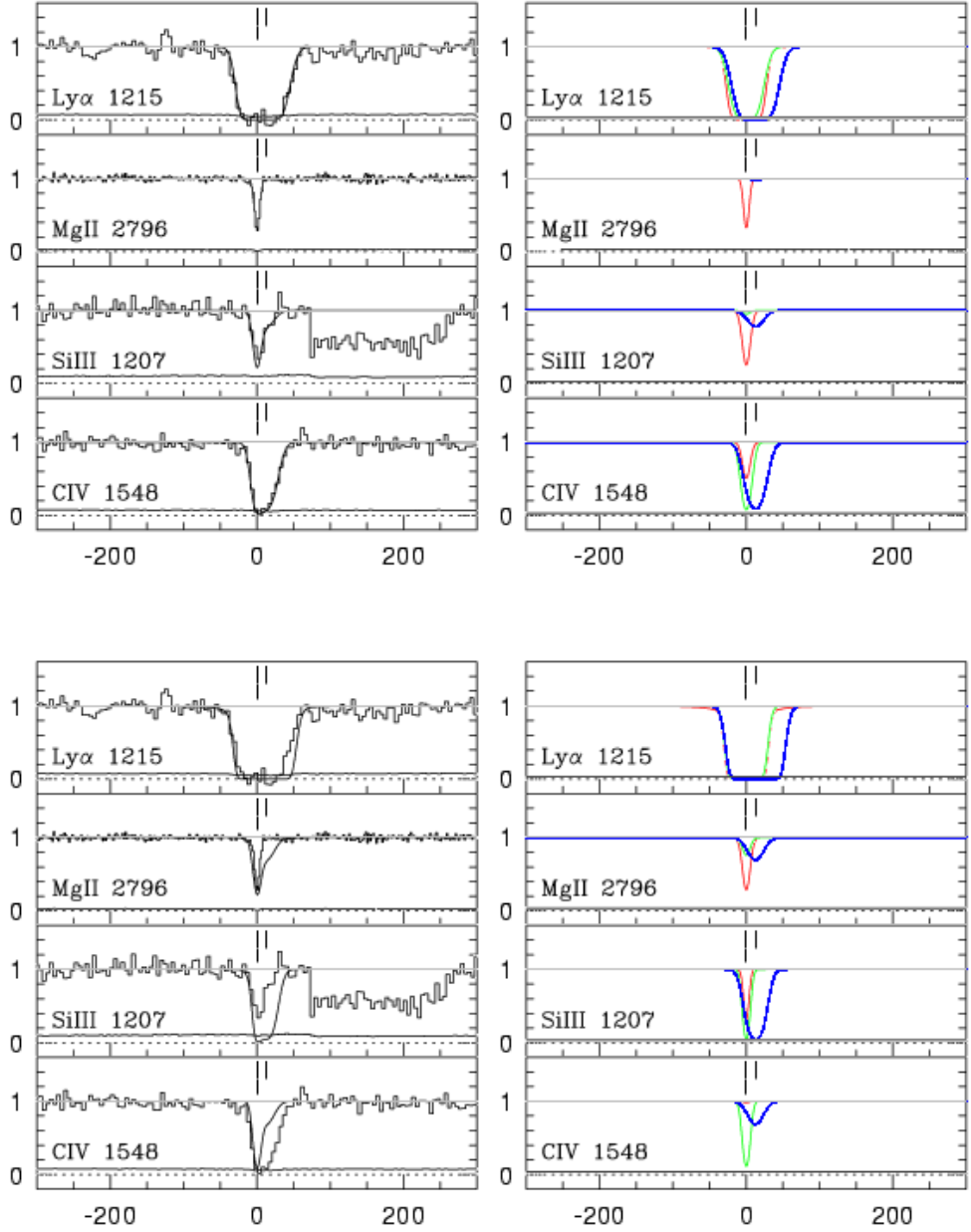}
\protect
\caption{\small{Evolution of the $z \sim$ 0.6 weak {\MgII} absorber from its observed redshift to the present epoch. {\bf Top Left Panel:} Data with photoionization model from Charlton {\etal}(2003) superimposed. {\bf Top Right Panel:} The four different component clouds with their contribution towards absorption shown separately. The tick marks placed above the features represent the center of the absorption feature for each component cloud. The {\it red} curve is the {\MgII} phase, centered at 0 km/s in velocity space. The other curves represent the three low--density, high ionization {\CIV} phases offset in velocity by 0 km/s, +24 km/s and +54 km/s. {\bf Bottom Left Panel:} Shows how the absorption feature of a similar system present at $z$ = 0 would look like. The data are superimposed on the synthetic spectrum to highlight changes in absorption feature strengths for individual transitions. {\bf Bottom Right Panel:}The four separate phases are shown for the evolved absorber at $z$ = 0. The changes in column density for {\MgII}, {\CIV} and {\HI} are summarized in Table~\ref{tab:tab7}}}
\label{fig:12}
\end{figure*}

For the purpose of this thought experiment, we assume that an absorber
exists at $z\sim0$ for which the total hydrogen column densities,
$N_{tot}$, electron number densities, $n_e$, and metallicities, $\log
Z$ (in solar units), of each of the two phases have the same values as
they have for the absorber observed at intermediate redshift.  These
values, listed in Table~\ref{tab:tab4}, are taken from \citet{weak1634}. In other words, we
are considering the effect of the EBR change on an otherwise static
structure.  The changing photon number density $n_{\gamma}$ from the
EBR leads to a change in the ionization parameter, $\log U =
n_{\gamma}/n_e$, which directly affects the ionization state of the
absorber.  The Cloudy code
\citep{cloudy01} was used to solve for the ionization state of each
phase of the absorber.  The changes in the column densities of the
various key transitions from the observed redshift to $z=0$ are listed
in Tables~\ref{tab:tab5}, \ref{tab:tab6}, and \ref{tab:tab7}.  We
then synthesized model spectra, noise--free, for the evolved model
systems at $z=0$ as they would appear if observed at a resolution of
$R=45,000$.  In order to synthesize these spectra, we used the
column densities and temperature output by Cloudy.  The latter was
used along with the turbulent $b$ parameter from the intermediate
redshift model cloud in order to determine the observed $b(H)$ and
the corresponding $b$ values for all elements. Figures~\ref{fig:10}, 
\ref{fig:11}, and \ref{fig:12} present
the results of this evolution experiment.  The individual model
components are shown separately at the observed redshift and at
$z=0$, as well as the result of combining these components.

The $z=0.81$ system, summarized in Table~\ref{tab:tab5} and
Figure~\ref{fig:10}, is the simplest system, with only a
single--component required to fit the high ionization phase
absorption.  In this case, at $z=0$ the $N({\MgII})$ contribution from
the high--ionization phase has increased by an order of magnitude,
while its $N({\CIV})$ has decreased by about the same factor.  This
leads to a substantially stronger weak, low--ionization absorber at
$z=0$, which has {\CIV} absorption that may not even be detected
in some spectra.

Table~\ref{tab:tab6} and Figure~\ref{fig:11} present results
for evolution of the $z=0.90$ system, which has two high--ionization
phase clouds, one aligned with the low--ionization absorption, and
one offset by $\sim15$~{\kms}.  As we expect, the {\MgII} absorption
strength increases and the {\CIV} absorption strength decreases at
$z=0$.  However, the most noteworthy change is that the {\MgII} absorption
from the offset high--ionization component becomes detectable at low
redshift.  This system may at $z=0$ be classified as a multiple cloud,
weak {\MgII} absorber.

The $z=0.65$ system, presented in Table~\ref{tab:tab7} and
Figure~\ref{fig:12}, would also evolve into a multiple--cloud,
weak {\MgII} absorber at $z=0$.  For this system, although the
{\CIV}~1548 absorption is substantially weaker at $z=0$ than at
$z=0.65$, it is still comparable to the {\MgII}~2796.

In general, the decreasing EBR will lead to an increase in the
{\MgII} equivalent width for a given system.  For these representative
systems, this does not occur as a result of an increased equivalent
width for the small, $\sim 1$--$100$~pc, structure that produces the
{\MgII} absorption at $z\sim1$.  Instead, it is the result of a large
increase in the {\MgII} absorption contributed by the larger
(hundreds of pcs to a few kpcs) structure that produced the {\CIV}
absorption at $z\sim1$.  Often the {\CIV} profiles for weak
{\MgII} absorbers at $z\sim1$ have multiple components \citep{weak1634}.
There could be detectable {\MgII} absorption at $z\sim0$ from each
of the structures that produces such a {\CIV} component at $z\sim1$.
Therefore, we expect a larger number of multiple--cloud, weak low
ionization absorbers at low redshift.  Also, there will be an
increase in $W_r(2796)$ centered on the original weak {\MgII}
absorber, both from the small, higher density structure and from
the large, lower density structure.  These effects will lead to
detection of a larger number of weak, low--ionization systems at
$z\sim0$ than at $z\sim1$.  An excellent example of this is the
$z=0.65$ system toward PG~$1634+459$ (see Figure~\ref{fig:12}).
In fact, this system was just below the detection threshold of
the \citet{weak1} survey and was not detected in that survey.
It would very easily be detectable if the same structure
existed at $z=0$, primary due to the {\MgII} contributed by
the larger, lower density phase clouds.

\subsection{Expected $dN/dz$ at $<z>=0.15$ Due to Evolving EBR}
\label{sec:4.3}

Assuming the observed equivalent width distribution for $<z>=0.9$,
we can estimate the expected $dN/dz$ for that same population,
evolved to $<z>=0.15$ (the median for our survey) subject to
the changing EBR.
Our survey was limited to weak systems corresponding to the
equivalent width interval $0.02 < W_r(2796) <0.3$~{\AA}.
We demonstrated in \S~\ref{sec:4.2} that at
$<z>=0.15$ these can result from the evolved {\MgII} phase
or the evolved {\CIV}
phase of an $<z>=0.9$ absorber.  They can also result
from $<z>=0.9$ {\CIV} clouds that are not related to detected {\MgII}
absorbers at that time.  For both of these possible predecessors
of the $<z>=0.15$ weak {\MgII} absorber population, we estimate
the expected $dN/dz$ result.
We calculate what interval of equivalent width of the {\MgII}
or {\CIV} cloud predecessors would give rise to $0.02 \leq W_r(2796) < 0.3$~{\AA}
at $<z>=0.15$.  Then we integrate the appropriate observed 
equivalent width distribution for {\MgII} or {\CIV} absorbers
at $z\sim1$ \citep{weak1,ss88,tripp96} using those new corresponding values
as limits.  This is an estimate of how many absorbers should be in
the observed interval at present.

First, we consider the $<z>=0.15$ weak {\MgII} absorbers that evolve
from {\MgII} clouds with $W_r(2796)<0.02$~{\AA} at $<z>=0.9$.
On the linear part of the {\MgII} curve of growth,
$W_r(2796)=0.02$~{\AA} corresponds to $N({\MgII}) \sim 10^{11.7}$~{\cm2}
for the full range of plausible Doppler $b$ parameters. For a
metallicity and density that is typical for weak
{\MgII} absorbers ($\log Z = -1$ in solar units and $n_H\sim10^{-2}$~{\cc}
\citep{weak1,weak1634}), a {\MgII} absorber
with $W_r(2796) = 0.02$~{\AA} at $<z>=0.15$ can arise from a single
cloud absorption system with total neutral hydrogen column density
$N_{Htot} \sim 10^{17.35}$~{\cm2}.
Under the influence of the stronger EBR at $z$ = 0.9,
a cloud structure with the same $N_{Htot}$, and with the same
metallicity, density, and size would produce an absorption feature
with $N({\MgII}) \sim 10^{11.34}$~{\cm2}.  This corresponds to
$W_r(2796)=0.008$~{\AA}.  We obtain the same number for a ``QSO--only'' EBR.

Next we consider how the strongest cloud in our survey [$W_r(2796)=0.3$~{\AA}
at $<z>=0.15$] would relate to the equivalent width distribution
at $<z>=0.9$. A single, low ionization cloud with
$N({\MgII}) =10^{15.8}$~{\cm2} will have $W_r(2796)=0.3$~{\AA} for
a typical Doppler $b$ parameter of $4$~{\kms}.  Assuming again the
typical parameters, $\log Z=-1$ and $n_H\sim10^{-2}$~{\cc}, we find that the
structure have $N_{Htot} = 10^{21.25}$~{\cm2}.  From that same structure,
at $<z>=0.9$, the {\MgII} absorption lines would be only slightly weaker,
with $W_r(2796)=0.29$~{\AA}.  In reality, it is more likely that a $W_r(2796)=0.3$~{\AA} absorber is produced
by multiple clouds.  If these are blended and saturated, the change
in $W_r(2796)$ would be even smaller.

We therefore expect that the physical structures that would produce
weak {\MgII} absorption at $z$ = 0.15 with $0.02 \leq W_r(2796) <
0.3$~{\AA}, would correspond to the equivalent width
interval $0.008 \leq W_r(2796) < 0.29$~{\AA} at $<z>=0.9$.
From this result and the assumption that weak {\MgII} absorbers
have similar total hydrogen column densities, densities, and
metallicities at $<z>=0.9$ and $<z>=0.15$, a value for $dN/dz$ 
that we would expect to observe at $<z>=0.15$ can be calculated
as follows.

The equivalent width distribution function, $n(W_r)$ gives the number of
{\MgII} absorption systems with rest frame equivalent width $W_r$ per unit
equivalent width per unit redshift path. The function is represented by the
power law expression
\begin{equation}
n(W_r)d(W_r) = CW_r^{-\delta}dW_r
\end{equation}
\noindent with $C\sim 0.4$ and $\delta = 1.04$ \citep{weak1}.
The ratio
\begin{equation}
\frac {\int_{0.008}^{0.29}n(W_r)dW_r}{\int_{0.02}^{0.3}n(W_r)dW_r} = 1.35
\end{equation}
\noindent shows the factor by which the number of absorbers
in the equivalent width interval of $0.008 \leq W_r(2796) < 0.29$~{\AA}
exceeds the number of absorbers in the interval $0.02 < W_r(2796) <
0.3$~{\AA}.  Therefore, the expected value for $dN/dz$ of weak {\MgII}
systems at the present epoch, leading from the evolution of the
{\MgII} phase of the absorber population at $z \sim 1$ is estimated to
be a factor of $\sim1.35$ higher than the statistical redshift number
density computed from the survey by \citet{weak1}.  Taking into account
the expected cosmological evolution, for a $\Omega_{\Lambda}=0.7$ cosmology (a factor
of $0.56$ decrease from $z=0.9$ to $z=0.15$), we
predict a contribution of $dN/dz = 1.3$ to the expected numbers of weak {\MgII} absorbers.

We consider how this estimate varies for different
parameters describing the $z=1$ population.  Based on previous
studies \citep{weak2,weak1634}, we consider $-1.5 < \log Z < 0$ and
$10^{-3} < n_H < 10^{-1}$~{\cc}, and $b$ values from $2$ to $8$~{\kms}.
In the optically thin regime, metallicity has little effect.
However, density has a strong effect on the change of $N({\MgII})$
with redshift.  This is due to the relatively strong dependence of
$N({\MgII})$ on the ionization parameter, particularly in the range
$-2.5 < \log U < -1.5$.  The photon number density $n_{\gamma}$ changes
from $-5.6$ to $-6.3$ over the range $0.9 \geq z \geq 0.15$.  
$\log U = \log n_{\gamma} - \log n_e$ is $-1.5$ at $<z>=0.15$ for a density of
$\log n_{\gamma}=-3.8$ [{\cc}].  Low densities produce the largest changes in
$N({\MgII})$.  Since the {\MgII} phases of the weak {\MgII} absorbers at $<z>=0.9$ have
higher densities, they are not very strongly affected by the EBR.
However, we have found in \S~\ref{sec:4.2} that the {\CIV}
phases of these absorbers do have a drastic change in their {\MgII}
column densities.  That is because their densities are in the range
$10^{-4} < n_e < 10^{-3}$~{\cc} \citep{weak1634}.

Next, we give a rough estimate of the $dN/dz$ of the population of weak {\MgII} absorbers
at $<z>=0.15$ that should result from the evolved population of {\CIV}
absorbers from $z\sim1$.  This is quite uncertain because of sensitivity to
uncertain physical parameters of {\CIV} absorbers, and because of uncertainties
in the observed equivalent width distribution of that population at $z\sim1$.
The equivalent width distribution for {\CIV}
absorbers is best measured for $z>1.3$ for which {\CIV} is in the optical.
We therefore estimate the equivalent width of the {\CIV} absorber,
$W_r(1548)$, at $z=1.3$ that would evolve to have $W_r(2796)=0.02$~{\AA}
at $z=0.15$.  For cloud densities $\log n_e=-4$ or $-3$~{\cc},
an absorber with $N_{Htot}=10^{19.5}$~{\cm2} or $10^{18.0}$~{\cm2}
would have $N({\MgII})=10^{11.7}$~{\cm2} [corresponding to $W_r(2796) = 0.02$~{\AA}].
A structure with these same $N_{Htot}$ values would have
$W_r(1548) =0.12$ or $0.03$~{\AA}, respectively, for the two densities,
at $z=1.3$.  Assuming a QSO--only EBR would decrease these values
somewhat [e.g. to $W_r(1548)=0.02$~{\AA} for $\log n_e=-3$~{\cc}], but
this change is not substantial compared to other uncertainties in our
estimate.

The next step is to determine the $dN/dz$ of {\CIV} absorbers with
$W_r(1548)>0.12$~{\AA} or $>0.03$~{\AA} at $z=1.3$.  One option is to
integrate the observed equivalent width distribution for {\CIV}
systems at $z=1.3$, extrapolated to lower $W_r(1548)$:
\begin{equation}
n(W_r) dW_r = (N_*/W_*) \exp(-W_r/W_*) dW_r
\end{equation}
\noindent with $N_*=4.60$ and $W_* = 0.46$~{\AA} down to $W_r(1548)=0.15$~{\AA} (sample A4 of \citet{ss88}).
We obtain $dN/dz = 3.5$, integrating from $W_r(1548)=0.12$~{\AA} to
$\infty$, corresponding to $n_e=10^{-4}$~{\cc}, and $dN/dz=4.3$,
integrating from $W_r(1548)=0.03$~{\AA} to $\infty$, corresponding to
$n_e=10^{-3}$~{\cc}.  
Even though
the largest change in $N({\MgII})$ is for $n_e \sim 10^{-3.8}$~{\cc}, we
find the larger $dN/dz$ for $n_e \sim 10^{-3}$~{\cc} since the fraction of
magnesium in the form of {\MgII} is larger at any redshift for such a
cloud.  
However, \citet{tripp96} produced a more sensitive and direct survey for
{\CIV} absorbers, particularly in the latter equivalent width range, $W_r(1548)>0.03$~{\AA}
at $1.5 < z < 2.9$.  They found a somewhat larger number, $dN/dz = 7.1 \pm 1.7$, though
it is consistent within $2\sigma$.  We estimate that $4 < dN/dz < 7$ for $z=1.3$ {\CIV}
absorbers that would evolve to have $W_r(2796)>0.02$~{\AA} at $<z>=0.15$.

Some of these absorbers would, however, already give rise to weak or
strong {\MgII} absorption at $<z>=1.3$. All but a small fraction of
{\MgII} absorbers, denoted as ``{\CIV}--deficient''
\citep{archive2}, do have {\CIV} detected at the same or similar velocity, though it is likely to
be in a different phase.  We estimate $dN/dz \sim 0.6$ at $<z>=1.3$
for $W_r(2796)>0.3$~{\AA}, using the redshift parameterization given
in \citet{nestor}.  Similarly, \citet{weak1} measured $dN/dz = 2.2 \pm
0.4$ at $1.07 < z < 1.4$.  Adding these two numbers, and subtracting
from the $dN/dz$ of the relevant {\CIV} population listed above yields
the range $1 < dN/dz < 4$ for $<z>=1.3$ {\CIV} absorbers that would
evolve into ``new'' weak {\MgII} absorbers at $<z>=0.15$.  We must
also again consider cosmological evolution of that population to
determine the expected $dN/dz$ of its low redshift counterparts.  For
an $\Omega_{\Lambda} = 0.7$ cosmology, $dN/dz$ changes by a factor of
$0.48$ from $<z>=1.3$ to $<z>=0.15$.  Therefore, $0.5 < dN/dz < 2$ is
our final estimate for the the expected number of {\MgII} absorbers
with $W_r(2796)>0.02$~{\AA} at $z=0.15$ due to the evolution of the
{\CIV} absorber population subject to the changing EBR.  We consider
the upper part of this range more probable since it was derived from
the more direct \citet{tripp96} measurement of the $z\sim1.3$ {\CIV} absorber population.

\subsection{Discussion of Comparison of Observed and Expected $dN/dz$
and the Nature of Weak {\MgII} Absorbers}
\label{sec:4.4}

We have found that the observed $dN/dz$ at $<z>=0.15$ is
$1.0\pm0.20$.  Considering sample bias, this is likely to be an upper
limit to the true number. However, taking two types of absorption
systems that exist at $z\sim1$ and evolving these populations to
$z=0.15$, we would expect a larger number, $2 < dN/dz < 3$.
This expectation includes the evolution of the {\MgII} phases of
the small, parsec--scale absorbers that produce weak {\MgII}
absorption at $z\sim1$ ($dN/dz\sim1$; hereafter {\it parsec--scale
structures}\footnote{Sizes of this phase ranges from $\sim1$~pc to
$\sim 100$~pc \citep{weak2, weak1634}}) and the evolution of the
{\CIV} absorbers at $z\sim1$ that would contribute to the weak {\MgII}
absorber population at lower redshift ($dN/dz\sim4$; hereafter {\it
kiloparsec--scale structures}).  Our expectations for $dN/dz$ at
$z=0.15$ resulting from absorber populations at $z\sim1$ is valid only
if the absorbers are physically stable over large time scales or if
their rate of formation per co--moving volume is constant.  Our
results suggest that one, or more likely both, of the types of
$z\sim1$ structures that evolve into low redshift weak {\MgII}
absorbers are evolving away.

It is important to note that, in fact, the expected increase in the
$dN/dz$ of weak MgII absorbers from $<z>=0.9$ or $<z>=1.3$ to lower
redshifts should already be apparent in the lowest redshift bin of the
\citet{weak1} survey, $<z>=0.57$.  If we repeat the estimate in
\S~\ref{sec:4.3} for $<z>=0.57$, we find that the expected increase
in the number of parsec--scale structures producing weak {\MgII}
absorption from $<z>=0.9$ is balanced almost exactly by cosmological
evolution, yielding an expected $dN/dz \sim 1.7$.  The expected
contribution to weak {\MgII} absorption from kiloparsec--scale
structures at $<z>=0.57$ is similarly estimated to be $dN/dz \sim 1$.
The combination is already twice the observed value in this redshift
bin of the \citet{weak1} survey.  Thus the indicated evolution is
occuring substantially from $z=1$ to $z=0.5$, not just at the
lower redshift range we have surveyed here.

In principle, the question of the predecessors of the $<z>=0.15$
population of weak {\MgII} absorbers is a simple one.  We should
simply examine the physical nature (e.g. density and size) of the
systems that produce the absorption at $<z>=0.15$ and see which
category they fall in (e.g. are they produced just by a kiloparsec--scale
structure or by a line of sight passing though both a parsec--scale and
a kiloparsec--scale structure).
However, in practice this is difficult to determine because of the
possibility of ``hidden phases''.  Specifically, there can be
narrow (a couple {\kms}), low ionization profiles superimposed on
a broader profile from a higher ionization phases.  As discussed
in \S~\ref{sec:4.3}, this is expected
to be more common at $<z>=0.15$ that at $<z>=0.9$ because of the
contribution to the low ionization absorption from the clouds that
produce {\CIV} absorption.  Figures~\ref{fig:10} -- \ref{fig:12}
show that at $<z>=0.15$, the three example systems toward
PG$1634+459$ have evolved so that, although present at the same
level as before, the parsec--scale, high density phase is not
likely to be distinguishable through profile fitting and photoionization
modeling.

We can begin to consider what type of weak {\MgII} absorbers
exist in the present by analyzing the six systems present in our
survey.  Several of the systems discussed in
\S~\ref{sec:3.1} have evidence for multiphase structure (the
$z=0.051$ system toward PG~$1211+143$, the $z=0.138$ system toward
PG~$1116+215$, the $z=0.167$ system toward PKS~$0405-123$, and the
$z=0.081$ system toward PHL~$1811$), but in several cases the
multiphase conditions are required by the presence of {\OVI} and not
specifically by {\CIV}.  The $z=0.0057$ absorber toward
RX~J$1230.8+0115$ and the $z=0.0053$ absorber toward 3C~273 could both
be single--phase \citep{rosenberg,tripp02}, but the former could be
produced by a kiloparsec--scale structure while the latter has a size of
only $70$~pc \citep{tripp02}.
It is also important to note that at least
two of the systems (the $z=0.051$ system toward RXJ 1230.8-115, the
$z=0.167$ system toward PKS 0405-123, and possibly the $z=0.138$ system toward
PG~$1116+215$) are multiple--cloud weak {\MgII} absorbers.
These could either arise from evolved multiple kiloparsec--scale cloud systems
or from several parsec--scale clouds in a larger structure.  It seems that
the $z\sim0$ population of weak {\MgII} absorbers is an inhomogeneous
group with multiple origins.  However, fundamentally, we cannot demonstrate
that any one of the systems definitely does not have a parsec--scale
structure along the line of sight.

Since individual cases cannot be classified, we turn to a statistical
comparison of the numbers of weak {\MgII} absorbers at $<z>=0.9$
and at $<z>=0.15$.  Roughly $dN/dz\sim1$ is observed at both
epochs, but we would expect $dN/dz\sim2$--$3$ at $<z>=0.15$ based upon
the evolving EBR.  This expectation was divided into $dN/dz\sim1$
from parsec--scale structures and $dN/dz\sim1$--$2$ from kiloparsec--scale
structures.  Based on these numbers, three possibilities exist: (1) virtually
all kiloparsec--scale structures evolve away from $<z>=0.9$ to $<z>=0.15$;
(2) most kiloparsec--scale structures evolve away {\it and} most parsec--scale
structures evolve away over this interval; (3) most kiloparsec--scale and
some fraction of parsec--scale structures evolve away.  Note, that the
fate of the parsec--scale structures is uncertain, but that
in all three possibilities either most or all of the kiloparsec--scale
structures evolve away.  With this in mind, we consider the expected
$dN/dz$ for different possible origins of each of these types of
structures.

The parsec--scale size and small velocity dispersions (few {\kms}) for the
structures responsible for the low ionization phase of weak {\MgII}
absorbers at $z\sim1$ suggests they might be unstable over
astronomical time scales. The clouds are not in simple pressure
equilibrium, since the temperatures for the high and low ionization
phases are approximately the same, with high density contrast. The
criterion for these gaseous regions to be confined by gravity can be
estimated from the method given in \citet{schaye}. For an optically
thin absorber to be gravitationally bound, the radial size of the
cloud is given by the expression

\begin{equation}
L \sim 10^2 \mathrm{kpc}
\left(\frac{N_{\HI}}{10^{14}cm^{-2}}\right)^{-\frac{1}{3}}T_{4}^{0.41}\Gamma_{12}^{-\frac{1}{3}}\left(\frac{f_{g}}{0.16}\right)^{\frac{2}{3}}
\end{equation}
\noindent where $N({\HI})$ is the column density
of neutral hydrogen for the absorbing region, $T$ the temperature in
units of Kelvin, $\Gamma_{12}$ is the hydrogen photoionization rate
normalized to 10$^{-12}~{\rm s}^{-1}$, and $f_g$ is the fraction of
mass in the form
of gas given by the ratio $\Omega_{b}$/$\Omega_{m}$. At low redshifts
the estimated rate of photoionization is $10^{-13}$ s$^{-1}$
\citep{DaveTripp2001}. The typical value of baryon mass gas fraction is
$\sim 0.16$ (\citet{schaye} and references therein). Hence, the size of
low ionization clouds which produce weak {\MgII} absorption must
exceed $\sim 10$--$20$~kpc in order to be gravitationally
stable, much larger than the parsec--scale sizes inferred from
photoionization models.

These structures would not be gravitationally stable unless $f_g$ was
much larger.  However, a larger $f_g$ is feasible, either due to dark
matter or due to additional baryons.
\citet{weak2} suggested that the weak {\MgII}
absorbers could be remnant gas between the stars in a population of
intergalactic star cluster, where the star to gas ratio could
be $\sim 10^4$.  This is comparable to present limits on the amount
of gas that could be present in globular clusters.  
In an intergalactic star cluster model, the {\CIV} absorption seen
in $z\sim1$ weak {\MgII} absorbers would be produced by gas
ejected by supernovae from the cluster, which is housed in the
dark matter halo that surrounds it \citep{weak2}.  Both the
parsec--scale and the kiloparsec--scale components in such
a star cluster/dark matter halo model should be relatively stable.

Alternatively, the structures could be transient, but could be
regenerated at a rate consistent with the observed $dN/dz$.
One possibility of this type, also discussed by \citet{weak2}, is that weak
{\MgII} absorption arises in structures related to supernova remnants,
or to superbubbles or superwinds as favored by \citet{stocke04}.
In this scenario, {\CIV} absorption would arise in an interface
layer between the low ionization shell or sheet and a hotter interior.
These structures would not be stable on cosmological timescales, but
similar structures would be continuously forming at a rate
approximately proportional to the star formation rate.  Since
the intensity of the EBR is found to decrease since $z\sim1$
\citep{hm96,hmconf}, a scenario of this type would predict
a decline in the overall number of weak {\MgII} absorbers
from $z\sim1$ to $z\sim0$.  

A third possibility for the origin of weak {\MgII} absorbers is
that they arise in high velocity clouds (HVCs), such as those observed
around the Milky Way.  The observations of {\OVI} HVCs are consistent with
coherent motions of extended sheets of gas which cover more than $60$\%
of the sky \citep{sembachhvc}. The {\OVI} has been hypothesized to
arise in turbulent boundaries between warm/cold clouds and the hot
intragroup medium through which they are travelling
\citep{sembachhvc,fox}.  These large structures are likely to
have multiphase layers, giving rise to {\CIV} as well as
coincident lower ionization absorption from the warm/cold clouds, which has
been observed in most sightlines \citep{wakker01}.
Most schemes for the origin of HVCs (e.g. material falling
into the Local Group, material ejected from the Galaxy, or
tidal debris related to satellites) would lead to the prediction
of a decreasing $dN/dz$ from $z\sim1$ to $z\sim0$.

We must also consider the possible origins of the kiloparsec--scale
structures that could evolve to produce weak {\MgII} absorption at
$z\sim0$.  These structures are similar to the high--ionization
phases of weak {\MgII} absorbers at $z=1$, and should be closely
related to the {\CIV} absorber population.
The {\CIV} systems observed at low redshift have been found to
be closely related to galaxies.  Absorption in {\CIV} is detected
over most of a $100h^{-1}$~kpc radius region surrounding all
$L_{B*}$ galaxies, with the radius scaling at $L_B^{0.5}$
\citep{chenlanzetta01}.  Those authors favor accreting satellites
as the main mechanism for metal production at these large
distances, though they also consider galactic fountains.
Either of these models would predict a transient population
which would evolve away due to decreasing numbers of satellites
or star formation activity.

\section{CONCLUSIONS}
\label{sec:5}

We have presented the results of a survey of $20$ {\it HST}/STIS
E140M spectra in order to determine the redshift path density of weak
{\MgII} absorbers [$0.02 < W_r(2796) < 0.3$~{\AA}] at $<z>=0.15$.
The redshift path covered by our survey was $5.3$ at a completeness
limit of $26$\%.
We found $dN/dz =1.0 \pm 0.20$, which is consistent with no evolution
from $<z>=0.9$, where \citet{weak1} found $dN/dz=1.74 \pm 0.10$.
Our survey may be slightly biased toward lines of sight with weak
{\MgII} absorbers, thus the actual number may be somewhat smaller.

The apparent lack of evolution is deceptive because it suggests no
change in the population of objects producing the absorption.
However, the extragalactic background radiation intensity is known to
decrease by a factor of $\sim8$ from $<z>=0.9$ to $<z>=0.15$
\citep{hm96,hmconf}.  Therefore the ionization state of absorbers will
change.  We considered how the three single--cloud, weak {\MgII}
absorbers observed toward PG~$1634+706$ at $z=0.65$, $z=0.81$, and
$z=0.90$ would change if the same structures existed in the present.
In general, the low ionization absorption grows stronger as the high
ionization absorption becomes weaker.
We found that the large, low density clouds that produced {\CIV}
absorption at $z\sim1$ will produce observable, but weak, {\MgII} at
$z\sim0$.  Thus what was a single--cloud weak {\MgII} absorber
at $z\sim1$ can become a multiple--cloud weak {\MgII} absorber
at $z\sim0$.

Because of the decreasing ionization state of absorbers from $z\sim1$
to $z\sim0$, there would be {\MgII} absorbers that fell below a
$0.02$~{\AA} threshold at $z\sim1$, but became detectable at $z\sim1$.
Considering the expected evolution, and integrating the observed
$z\sim1$ equivalent width distribution, we found that in fact
this does not lead to a significant increase in the weak {\MgII}
absorber population at $z\sim0$.  The {\MgII} column density
is not very sensitive to ionization parameter in the relevant range.
We predict a $dN/dz\sim1$ for the $<z>=0.15$ population of parsec--scale,
weak {\MgII} absorbers.
On the other hand, there are also many kilparsec--scale {\CIV}
absorption systems that did not have detectable {\MgII} absorption at
$z\sim1$, but would become detectable in {\MgII} at $z\sim0$.
Again, considering the $z\sim1$ population of {\CIV} absorbers and
their properties, and assuming they are stable and change only
due to the changing EBR, we expect $dN/dz\sim1$--$2$ at $<z>=0.15$.

We conclude that the evolved population of structures observed to
exist at $z\sim1$ should give rise to roughly two to three times more weak
{\MgII} absorbers than are observed at $z\sim0$.  This expected
$z\sim0$ population would likely be dominated by kiloparsec--scale, low
density structures that only gave rise to {\CIV} absorption at
$z\sim1$.  However, it would also have a $30$--$50$\% 
contribution from the evolved 
population of parsec--scale, higher density absorbers that produced
weak {\MgII} absorption at $z\sim1$.  The numbers indicate that
most kiloparsec--scale structures are evolving away from $z\sim1$
until the present.  If all of them are evolving away, then it is
possible that none of the parsec--scale structures are evolving
away.  This could mean that they are stable or that they are
regenerated at a roughly constant rate.  On the other hand, it
seems somewhat more likely that some of the kiloparsec--scale structures will
remain at $z\sim0$.  They could be related to satellites and their
interactions or to superwind activity, effects that would decrease
with time, but would not disappear.  If so, then the parsec--scale
structures would have to be transient, and their rate of generation
would have to decrease with time.  This would be consistent with
models that explain weak {\MgII} absorption as supernova remnants,
supershells, superwinds, or high velocity clouds, but not necessarily
with intergalactic star clusters.

Although we have shown that weak {\MgII} absorbers exist at $z\sim0$,
we only now have hints about the relationship between the structures
that produce this population and those that produce similar {\MgII}
absorption at $z\sim1$.  If the low ionization absorption profiles
could be observed at much higher resolution ($R>100,000$), it should
be possible to dissect the $z\sim0$ population into parsec--scale and
kiloparsec--scale components.  Clearly, studies that identify luminous
objects that are related to the $z\sim0$ population are of great
value.  For both of those types of studies of $z\sim0$ weak {\MgII}
absorbers, a larger sample would be quite useful (larger than the six
mentioned in this paper).  For this, future UV spectroscopy missions are
essential.  Finally, charting the evolution of the weak
{\MgII} absorber population to redshifts of $z\sim2$ and $z\sim3$
will also provide constraints on its origins.

This research was funded by NASA under grants NAG~$5$--$6399$ and
NNG$04$GE$73$G and by NSF under grant AST--$04$--$07138$. JRM and
RSL were partially funded by the NSF REU program.


\newpage

\newpage

\newpage

\newpage

\newpage

\newpage

\newpage

\newpage

\newpage

\newpage

\newpage

\newpage


\begin{deluxetable}{llcccclc}
\tablenum{1}
\tabletypesize{\footnotesize}
\tablecaption{{\it STIS}/HST E140M Grating Data}
\tablehead{
\colhead{QSO ID} &
\colhead{$z_{\textrm{\scriptsize QSO}}$} &
\colhead{} &
\colhead{$S/N$ ratio$^{\dag}$} &
\colhead{} &
\colhead{$t_{exp}$} &
\colhead{PI} &
\colhead{Proposal ID} \\
\colhead{} &
\colhead{} &
\colhead{1216\AA} &
\colhead{1335\AA} &
\colhead{1548\AA} &
\colhead{kilosec} &
\colhead{} &
\colhead{}
}
\startdata
PKS 0405-123 & $0.534$ & $4.5$ & $6.4$ & $14.8$ & $27.2$ & Heap & $7576$ \\
PG 0953+415 & $0.239$ & $5.9$ & $8.3$ & $5.8$ & $24.5$ & Savage & $7747$ \\
PG 1116+215 & $0.177$ & $9.9$ & $10.9$ & $8.8$ & $39.8$ & Sembach/Jenkins & $8097/8165$ \\
3C273 & $0.158$ & $16.3$ & $24.4$ & $18.6$ & $18.7$ & Heap & $8017$ \\
RX J1230.8+0115 & $0.117$ & $5.4$ & $9.5$ & $6.3$ & $27.2$ & Rauch & $7737$ \\
PG 1259+593 & $0.478$ & $7.0$ & $9.0$ & $5.8$ & $95.8$ & Tripp & $8695$ \\
PKS 1302-102 & $0.278$ & $5.2$ & $8.2$ & $8.1$ & $22.1$ & Lemoine & $8306$ \\
3C351.0 & $0.371$ & $5.0$ & $9.2$ & $6.5$ & $74.5$ & Jenkins & $8015$ \\
H 1821+643 & $0.297$ & $8.1$ & $19.4$ & $11.5$ & $50.9$ & Jenkins & $8165$ \\
PKS 2155-304 & $0.116$ & $13.2$ & $15.9$ & $14.1$ & $28.4$ & Shull & $8125$ \\
PG 1211+143 & $0.081$ & $12.5$ & $24.8$ & $12.1$ & $14.3$ & Shull & $8571$ \\
HS 0624+6907 & $0.370$ & $5.0$ & $8.8$ & $7.1$ & $62.0$ & Tripp & $9184$ \\
3C249.1 & $0.311$ & $5.6$ & $8.4$ & $5.2$ & $68.8$ & Tripp & $4939$ \\
PG 1444+407 & $0.267$ & $4.1$ & $6.8$ & $7.9$ & $49.0$ & Tripp & $9184$ \\
HE 0226-4110 & $0.495$ & $5.6$ & $7.7$ & $7.9$ & $43.0$ & Tripp & $9184$ \\
PHL 1811 & $0.190$ & $6.4$ & $9.1$ & $7.1$ & $18.4$ & Jenkins & $9418$ \\
PKS 0312-77 & $0.223$ & $4.5$ & $6.3$ & $4.2$ & $37.9$ & Kobulnicky & $8312$ \\
TON S210 & $0.116$ & $4.0$ & $9.2$ & $5.0$ & $24.2$ & Sembach & $9415$ \\
PG 1216+069 & $0.331$ & $4.2$ & $7.2$ & $4.6$ & $69.8$ & Tripp & $9184$ \\
TON 28 & $0.330$ & $4.4$ & $7.0$ & $4.2$ & $48.4$ & Tripp & $9184$ \\
\hline
\enddata
\tablecomments{$\dag$ $S/N$ is per pixel}
\label{tab:tab1}
\end{deluxetable}
\clearpage

\newpage
\begin{deluxetable}{lcccc}
\tablenum{2}
\tablecaption{System Properties} 
\tablehead{
\colhead{z$_{absorber}$} & 
\colhead{QSO} & 
\colhead{W$_r$(1260) {\AA}} & 
\colhead{W$_r$(1335) {\AA}} & 
\colhead{Z(W$_r$,R)}
} 

\startdata
{0.005260}  & $3$C$273$ & $0.010{\pm}0.002$ & $0.010{\pm}0.002$ & $---$\\
{0.005671}  & RX~J$1230.8+0115$ & $0.10{\pm}0.01$ & $0.11{\pm}0.01$ & $5.28$\\
{0.051216}  & PG$1211+143$ & $0.020{\pm}0.005$ & $0.03{\pm}0.01$ & $4.90$\\
{0.080917}  & PHL $1811$ & $\sim 0.16^{\dag}$ & $0.16{\pm}0.01$  & $5.29$\\
{0.138489}  & PG$1116+215$ & $0.057{\pm}0.003$ & $0.082{\pm}0.005$ & $5.24$\\
{0.167121}  & PKS$0405-123$ & $0.13{\pm}0.01$ & $0.19{\pm}0.01$ & $5.29$\\

\hline
\enddata
\tablecomments{$\dag$ Obtained by fitting a Gaussian to data.}
\label{tab:tab2}
\end{deluxetable}
\clearpage

\newpage
\begin{deluxetable}{cccc}
\tablenum{3}
\tablewidth{0pt}
\tabletypesize{\footnotesize} 
\tablecaption{Results from Voigt Profile Fitting for Selected
Transitions} 
\tablehead{ 
\colhead{Transition} & 
\colhead{$v$ [{\kms}]} &
\colhead{$\log N$ atoms cm$^{-2}$} & 
\colhead{$b$ [{\kms}]}
} 
\startdata
\multicolumn{4}{c}{\sc 3C273, $z = 0.005260$ system}\\
\hline
{\SiII} & $0$ & $11.90\pm0.07$ & $9.4\pm2.0$ \\
{\CII} & $0$ & $12.79\pm0.08$ & $15.4\pm4.0$ \\
\hline
\multicolumn{4}{c}{\sc RX~J$1230.8+0115$, $z = 0.005671$ system}\\
\hline
{\SiII} & $0$ & $12.64\pm0.05$ & $6.8\pm1.0$ \\
{\SiII} & $-33$ & $12.56\pm0.07$ & $15.7\pm3.2$ \\
{\CII} & $3$ & $13.71\pm0.02$ & $12.2\pm0.7$ \\
{\CII} & $-33$ & $13.26\pm0.05$ & $8.9\pm1.2$ \\
{\CIV} & $1$ & $13.21\pm0.04$ & $7.1\pm0.9$ \\
\hline
\multicolumn{4}{c}{\sc PG 1211 + 143, $z=0.051216$ system}\\
\hline
{\SiII} & $0$ & $12.31\pm0.02$  & $5.3\pm0.5$ \\
{{\CII}$^{a}$} & $0$ & \nodata & \nodata \\
{\CIV} & $4$  & $13.95\pm0.02$  & $13.5\pm0.5$ \\
{\CIV} & $-23$  & $13.12\pm0.05$  & $6.8\pm1.0$ \\
{\CIV} & $-53$  & $13.06\pm0.07$  & $9.6\pm1.8$ \\
{\CIV} & $-77$  & $12.8\pm0.1$  & $12.2\pm4.3$ \\
\hline
\multicolumn{4}{c}{\sc PHL 1811, $z=0.080917$ system}\\
\hline
{{\SiII}$^{b}$} & $0$ & \nodata  & \nodata  \\
{\CII} & $0$ & $14.40\pm0.05$  & $11.4\pm0.5$ \\
{\CIV} & $7$  & $14.05\pm0.04$  & $11.4\pm0.7$ \\
{\CIV} & $-47$ & $13.54\pm0.05$  & $22.8\pm 3.6$  \\
\hline
\multicolumn{4}{c}{\sc PG 1116 + 215, $z=0.138489$ system}\\
\hline
{\SiII} & $0$ & $12.79\pm0.01$  & $7.2\pm0.2$ \\
{\CII} & $2$ & $13.87\pm0.03$ & $10.0\pm0.6$ \\
{\SiIV} & $-2$ & $12.60\pm0.06$  & $6.6\pm1.5$ \\
\hline
\multicolumn{4}{c}{\sc PKS 0405 - 123, $z = 0.167121$ system}\\
\hline
{\SiII} & $0$ & $13.25\pm0.05$ & $9.7\pm0.7$ \\
{\CII}  & $2$ & $13.73\pm0.06$ & $8.2\pm1.2$ \\
{\CII}  & $-35$ & $14.26\pm0.07$ & $11.6\pm1.0$ \\

\hline
\enddata
\tablecomments{Velocities are offsets from strongest {\SiII} $\lambda$1260{\AA} component.  (a) The {\CII} $\lambda$1335{\AA} feature is blended with Galactic {\SiIV} $\lambda$1403{\AA}.  (b) {\SiII} $\lambda$1260{\AA} absorption feature is blended, probably with {\Lya} at $z=0.1205$ (refer to Fig~\ref{fig:5}).  
}
\label{tab:tab3}
\end{deluxetable}
\clearpage

\newpage
\begin{deluxetable}{lrcclrccccccccc}
\tablenum{4}
\tabletypesize{\scriptsize} \rotate \tablewidth{0pt}
\tablecaption{Cloud Properties for the three single cloud weak
systems from the PG 1634 + 706's spectrum} \tablehead{ \colhead{} &
\colhead{$v$}  & \colhead{$Z$} & \colhead{$\log U$} &
\colhead{$n_H$} & \colhead{size} & \colhead{$T$} &
\colhead{$N_{\rm tot}({\rm H})$} & \colhead{$N({\HI})$} &
\colhead{$N({\MgII})$} & \colhead{$N({\SiIV})$} &
\colhead{$N({\CIV})$} & \colhead{$b({\rm H})$} & \colhead{$b({\rm
Mg})$} &
\colhead{$b({\rm C})$} \\
\colhead{} & \colhead{[{\kms}]} & \colhead{[$Z_{\odot}$]} &
\colhead{} & \colhead{[{\cc}]} & \colhead{[pc]} & \colhead{[K]} &
\colhead{[{\cmsq}]} & \colhead{[{\cmsq}]} & \colhead{[{\cmsq}]} &
\colhead{[{\cmsq}]} & \colhead{[{\cmsq}]} & \colhead{[{\kms}]} &
\colhead{[{\kms}]} & \colhead{[{\kms}]} }

\startdata
\multicolumn{15}{c}{\sc $z=0.8181$ System} \\
\hline
{\MgII}$_{\rm 1}$ & $0$  & $2.0$ & $-4.0$ & $0.06$ & $0.1$ & $4600$ &  $16.3$ & $15.3$ & $12.0$ & $9.7$  & $9.4$  &  $9$ & $2$  & $3$ \\
{\CIV}$_{\rm 1}$ & $0$   & $2.0$ & $-2.0$ & $0.0006$ & $80$ & $8000$ & $17.2$ & $14.1$ & $10.7$ & $12.3$ & $13.5$ & $15$ & $10$ & $10$ \\
\hline
\multicolumn{15}{c}{\sc $z=0.9055$ System} \\
\hline
{\MgII}$_{\rm 1}$ & $0$  & $0.0$ & $-2.7$ & $0.003$ & $150$ & $9000$ &  $18.1$ & $15.8$ & $12.5$ & $13.1$  & $13.6$  &  $12$ & $3$  & $4$ \\
{\CIV}$_{\rm 1}$ & $0$   & $0.0$ & $-1.5$ & $0.0002$ & $1500$ & $14000$ & $18.0$ & $14.2$ & $9.1$ & $11.4$ & $14.0$ & $17$ & $4$ & $6$ \\
{\CIV}$_{\rm 2}$ & $15$   & $0.0$ & $-1.8$ & $0.0004$ & $600$ & $14000$ & $17.9$ & $14.4$ & $10.2$ & $12.2$ & $14.0$ & $20$ & $14$ & $14$ \\
\hline
\multicolumn{15}{c}{\sc $z=0.6534$ System} \\
\hline
{\MgII}$_{\rm 1}$ & $0$  & $0.1$ & $-4.0$ & $0.06$ & $2$    & $11000$ &  $17.5$ & $16.3$ & $11.8$ & $9.9$  & $9.6$  &  $14$ & $4$  & $5$ \\
{\CIV}$_{\rm 1}$ & $0$   & $0.1$ & $-2.5$ & $0.002$ & $1500$ & $19000$ & $19.0$ & $16.1$ & $11.8$ & $12.8$ & $13.7$ & $21$ & $12$ & $13$ \\
{\CIV}$_{\rm 2}$ & $24$   & $0.1$ & $-2.2$ & $0.001$ & $3200$ & $22000$ & $19.0$ & $15.8$ & $11.2$ & $12.6$ & $13.9$ & $20$ & $8$ & $9$ \\
{\CIV}$_{\rm 3}$ & $54$   & $0.1$ & $-2.2$ & $0.001$ & $2500$ & $22000$ & $18.9$ & $15.7$ & $11.1$ & $12.5$ & $13.8$ & $23$ & $14$ & $14$ \\

\hline
\enddata
\tablecomments{ \baselineskip=0.7\baselineskip Column densities
are listed in logarithmic units. \newline
 This table of values is taken from Charlton {\etal}2003. The three absorbers were studied using a combination of STIS/HST and KECK/HIRES spectra.}
\label{tab:tab4}
\end{deluxetable}
\clearpage

\newpage
\begin{deluxetable}{lcc}
\tablenum{5}
\tablecaption{Change in Column Density Due to Declining Ionising Background \newline $z=0.8$ to $z=0$}
\tablewidth{0pt}
\tablehead{
\colhead{Species} & \colhead{{\MgII} phase} & \colhead{{\CIV} phase} \\
\colhead{} & \colhead{}	& \colhead{$v$ = 0 km/s} \\
\colhead{} & \colhead{$log N$} & \colhead{$log N$}
}

\startdata
$N({\MgII})$ & $12.0 \mapsto 12.2$ & $11.2 \mapsto 12.4$ \\
$N({\CIV})$ & $9.2 \mapsto 7.6$ & $13.5 \mapsto 12.5$ \\
$N({\HI})$ & $15.3 \mapsto 16.0$ & $14.3 \mapsto 15.3$ \\
$N({\HII})$ & $16.3 \mapsto 16.0$ & $17.4 \mapsto 17.4$ \\
\enddata
\medskip
\tablecomments{Number density of Lyman limit photons from the Haardt--Madau spectrum with normalization log(n$_{\gamma}$) = -5.51 at $z=0.8$ and log(n$_{\gamma}$) = -6.48 at $z=0$} 
\label{tab:tab5}
\end{deluxetable}
\clearpage

\newpage
\begin{deluxetable}{lccc}
\tablenum{6}
\tablecaption{Change in Column Density Due to Declining Ionising Background \newline $z=0.9$ to $z=0$}
\tablewidth{0pt}
\tablehead{
\colhead{Species} & \colhead{{\MgII} phase} & \colhead{{\CIV} phase} & \colhead{{\CIV} phase} \\
\colhead{}        & \colhead{}		& \colhead{$v$ = 0 km/s} & \colhead{$v$ = 15 km/s} \\
\colhead{}	  & \colhead{$log N$}	& \colhead{$log N$}     & \colhead{$log N$}
}

\startdata
$N({\MgII})$  & $12.5 \mapsto 13.2$ & $9.8 \mapsto 11.9$ & $10.9 \mapsto 12.3$ \\
$N({\CIV})$   & $13.1 \mapsto 11.4$ & $14.0 \mapsto 13.5$ & $13.9 \mapsto 13.1$ \\
$N({\HI})$  & $15.5 \mapsto 16.6$ & $14.2 \mapsto 15.3$  & $14.5 \mapsto 15.6$  \\
$N({\HII})$  & $18.0 \mapsto 17.9$ & $17.9 \mapsto 19.1$ & $17.9 \mapsto 17.9$\\
\enddata
\tablecomments{Number density of Lyman limit photons from the Haardt--Madau spectrum with normalization log(n$_{\gamma}$) = -5.43 at $z=0.9$ and log(n$_{\gamma}$) = -6.48 at $z=0$}
\label{tab:tab6}
\end{deluxetable}
\clearpage

\newpage
\begin{deluxetable}{lcccc}
\tablenum{7}
\tablecaption{Change in Column Density Due to Declining Ionising Background \newline $z=0.6$ to $z=0$}
\tablewidth{0pt}
\tablehead{
\colhead{Species} & \colhead{{\MgII} phase} & \colhead{{\CIV} phase} & \colhead{{\CIV} phase} & \colhead {{\CIV} phase} \\
\colhead{}	& \colhead{}	& \colhead{$v$ = 0 km/s} & \colhead{$v$ = 24 km/s} & \colhead{$v$ = 54 km/s} \\
\colhead{}	& \colhead{$log N$}	& \colhead{$log N$}  & \colhead{$log N$}  & \colhead{$log N$}}
\startdata
$N({\MgII})$ & $11.8 \mapsto 12.0$ & $11.9 \mapsto 12.8$ & $11.1 \mapsto 12.3$ & $10.9 \mapsto 12.1$ \\
$N({\CIV})$ & $9.4 \mapsto 7.9$ & $13.7 \mapsto 12.8$ & $13.9 \mapsto 13.4$ & $13.8 \mapsto 13.3$ \\
$N({\HI})$ & $16.2 \mapsto 16.8$ & $16.1 \mapsto 16.9$ & $15.7 \mapsto 16.5$ & $15.5 \mapsto 16.4$ \\
$N({\HII})$ & $17.5 \mapsto 17.4$ & $19.1 \mapsto 19.1$ & $19.1 \mapsto 19.1$ & $19.0 \mapsto 18.5$ \\
\enddata
\tablecomments{Number density of Lyman limit photons from the Haardt--Madau spectrum with normalization log(n$_{\gamma}$) = -5.67 at $z=0.6$ and log(n$_{\gamma}$) = -6.48 at $z=0$ }
\label{tab:tab7}
\end{deluxetable}
\clearpage

\end{document}